\documentclass[times]{fldauth}

\usepackage{moreverb}

\usepackage[colorlinks,bookmarksopen,bookmarksnumbered,citecolor=red,urlcolor=red]{hyperref}

\newcommand\BibTeX{{\rmfamily B\kern-.05em \textsc{i\kern-.025em b}\kern-.08em
T\kern-.1667em\lower.7ex\hbox{E}\kern-.125emX}}

\def\volumeyear{2013}

\usepackage{subfig}
\usepackage{ mathrsfs } 
\DeclareMathAlphabet{\mathscrbf}{OMS}{mdugm}{b}{n}

\newcommand{\vecr}[1]{\mathbf{#1}}
\newcommand{\matx}[1]{\mathbf{#1}}
\newcommand{\vecrsym}[1]{\boldsymbol{#1}}

\begin{document}

\runningheads{J.~Alexandersen et al.}{Topology optimisation for natural convection problems}

\title{Topology optimisation for natural convection problems}

\author{Joe Alexandersen\corrauth\footnotemark[2], Niels Aage, Casper Schousboe Andreasen and Ole Sigmund}

\address{Department of Mechanical Engineering, Solid Mechanics, Nils Koppels All\'e, Technical University of Denmark, DK-2800 Kgs. Lyngby, Denmark}

\corraddr{Joe Alexandersen, Department of Mechanical Engineering, Solid Mechanics, Nils Koppels All\'e, Technical University of Denmark, DK-2800 Kgs. Lyngby, Denmark}

\cgs{Villum Fonden - NextTop project \\ The Danish Center for Scientific Computing (DCSC) }

\footnotetext[2]{E-mail: joealex@mek.dtu.dk}

\begin{abstract}
This paper demonstrates the application of the density-based topology optimisation approach for the design of heat sinks and micropumps based on natural convection effects. The problems are modelled under the assumptions of steady-state laminar flow using the incompressible Navier-Stokes equations coupled to the convection-diffusion equation through the Boussinesq approximation. In order to facilitate topology optimisation, the Brinkman approach is taken to penalise velocities inside the solid domain and the effective thermal conductivity is interpolated in order to accommodate differences in thermal conductivity of the solid and fluid phases. The governing equations are discretised using stabilised finite elements and topology optimisation is performed for two different problems using discrete adjoint sensitivity analysis. The study shows that topology optimisation is a viable approach for designing heat sink geometries cooled by natural convection and micropumps powered by natural convection.
\end{abstract}


\maketitle

\vspace{-6pt}

\section{Introduction} \label{sec:intro}
\vspace{-2pt}

Natural convection is an interesting and important phenomenon where fluid motion is induced due to spatial differences in the buoyancy force. These differences can be due to variations in concentration or temperature, among others. The most often considered type of natural convection is that due to temperature differences leading to variations in the fluid density, which is the type treated in this paper. Natural convection is a strongly coupled phenomenon where the temperature field induces fluid motion, which affects the temperature field through the convection of heat. Natural convection can thus be exploited for either the convective cooling effect, as is the case in e.g. electronics cooling \cite{Bar-Cohen2003}, or the fluid motion induced through differences in buoyancy \cite{Krishnan2004}.

Structural optimisation is the classical engineering discipline of modifying the design of a structure in order to improve its performance with respect to some desirable behaviour.
Simple, yet effective, structural optimisation techniques, such as size and configuration optimisation, are frequently applied to the design of heat sinks in electronics cooling. For instance,
Morrison \cite{Morrison1992} optimises plate fin heat sinks in natural convection using a downhill simplex method and empirical correlations. Morrison considers the fin thickness, fin spacing and backplate thickness as design variables.
Bahadur and Bar-Cohen \cite{Bahadur2005} optimise staggered pin fin heat sinks for natural convection cooled microprocessor applications using analytical equations. Here the design variables considered are pin height, diameter and spacing.
Simple forms of optimisation have likewise been applied to the design of buoyancy-driven PCR-reactors (polymerase chain reaction) by e.g. changing the aspect-ratios of reactor cylinders \cite{Muddu2011} or the diameter, length and configuration of closed-looped channel systems \cite{Krishnan2004}.

While these traditional optimisation techniques can provide significant improvements to existing designs, they are all limited in the design freedom as an a priori determined initial design must be supplied. This is where the topology optimisation method triumphs by having the possibility to find unintuitive and unanticipated designs.
Topology optimisation as it is known today was pioneered by Bends\o e and Kikuchi \cite{Bendsoee1988} as a material distribution method for finding an optimal structural layout, for a given problem subject to design constraints. The most popular numerical method for topology optimisation, now known as the density or SIMP (solid isotropic microstructure with penalisation) approach, was developed concurrently to the homogenisation approach \cite{Bendsoee1988}. The SIMP approach was originally suggested by Bends\o e \cite{Bendsoee1989} and used extensively by Zhou, Rozvany and coworkers \cite{Zhou1991,Rozvany1992}.
Although the topology optimisation method originated and gained maturity within structural mechanics, the method has since been extended to a wide range of physics, such as acoustics \cite{Duhring2008}, photonics \cite{Jensen2011}, fluidics \cite{Borrvall2003} and many more.

One of the extensions of the density-based topology optimisation method has been to purely conductive heat transfer \cite{Bendsoee2003,Gersborg-Hansen2006}. In order to take the heat transfer to an ambient fluid into account in the design process, a constant out-of-plane convection coefficient has been applied in many works for two-dimensional problems, e.g. \cite{Sigmund2001a}. For plane problems where the out-of-plane dimension is small, the in-plane convection is therefore neglected. However, in order to include the design-dependent effects of in-plane convective heat transfer, a common extension is to introduce some form of interpolation of the convection boundaries\footnote{This is necessary in order to treat fully three-dimensional problems, but this has to the authors' knowledge not been demonstrated in the literature.}.
Yin and Anathasuresh \cite{Yin2002} used a density-based peak interpolation function, Yoon and Kim \cite{Yoon2005} introduced a special type of parameterised connectivity between elements, Bruns \cite{Bruns2007} suggested to interpolate the convection coefficient as a function of the density-gradient and Iga et al. \cite{Iga2009} used a density-based smeared-out Hat-function, which also tried to take variation in the strength of the convective heat transfer into account.
Another approach, has been to track the boundary implicitly using the levelset approach to topology optimisation \cite{Ahn2010a}.

A common feature of the above works, except reference \cite{Iga2009}, is that while the dependency of the convective heat transfer on the exposed surface area is taken into account, a single constant convection coefficient, $h$, is assumed and applied on all convection boundaries. It is standard engineering practice to assume an average/effective convection coefficient and apply it in analysis, but these values are often taken from tables of empirical data or empirical models for very specific types of problems. During the topology optimisation process, the design changes and thereby the interaction with the ambient fluid changes. This, along with the fact that topology optimisation often leads to unanticipated designs, makes it hard to justify the application of a predetermined and constant convection coefficient based on empirical assumptions. This is one of the reasons for why it is necessary to extend topology optimisation to problems where both the solid and the ambient fluid is modelled, as will be discussed below.

Topology optimisation for fluid flow problems was pioneered for Stokes flow by Borrvall and Petersson \cite{Borrvall2003}. They achieved control of the topology of a solid domain in Stokes flow by the introduction of a friction term based on lubrication theory, yielding the generalised Stokes equations. The same methodology was later extended to the Navier-Stokes equations \cite{Gersborg-Hansen2005,Olesen2006}. The friction term has later become somewhat decoupled from lubrication theory and now constitutes the Brinkman approach to fluid topology optimisation, where the friction term can be seen as that arising from the introduction of an idealised porous medium.
The Brinkman approach has since been used for transport problems \cite{Andreasen2009}, reactive flows \cite{Okkels2007}, transient flows \cite{Deng2011,Kreissl2011}, fluid-structure interaction \cite{Yoon2010a} and flows driven by constant body forces \cite{Deng2012a}.
Alternatives to Brinkman penalisation exists in the literature: Guest and Prevost \cite{Guest2006} utilised the interpolation between two physical models, namely the Darcy and Stokes equations, and the levelset approach to topology optimisation has also been applied to fluid flow problems \cite{Zhou2008,Challis2009}, recently also in combination with the extended finite element method (X-FEM) \cite{Kreissl2012}.

With the possibility of performing topology optimisation of heat conduction problems as well as fluid problems, the logical next step is coupling these to perform topology optimisation of multiphysics convection-diffusion problems.
The difference in thermal conductivity of the solid and fluid domains is extremely important to take into account when dealing with problems where the temperature field inside the solid is of interest, which is the case for e.g. heat sinks. Many papers either concentrate on the temperature distribution of the fluid itself, e.g. \cite{Andreasen2009}, ignore the temperature problem in the solid domain \cite{Kontoleontos2012} or ignore the differences in thermal conductivity, e.g. \cite{Matsumori2013}. However, there are several notable papers taking the difference into account.
Yoon \cite{Yoon2010} interpolates the conductivity and other parameters in order to design heat dissipating structures subjected to forced convection. 
Dede \cite{Dede2010} presented results for jet impingement surface cooling problems using linear interpolation for the thermal conductivity.
Lee \cite{Lee2012} interpolates the conductivity and presents many interesting results for the design of convective cooling systems. 
McConnel and Pingen \cite{McConnell2012} interpolates the thermal diffusivity for the design of layered pseudo-3D problems using the Lattice-Boltzmann-Method.
Marck et al. \cite{Marck2013} presented topology optimisation of multi-objective heat exchanger problems using the finite volume method.
Lastly, Koga et al. \cite{Koga2013} recently presented optimisation and experimental results for a water-cooled device for compact electronic components.

\begin{figure}[b]
\centering
\subfloat[Forced convection]{\includegraphics[height=0.3\textwidth]{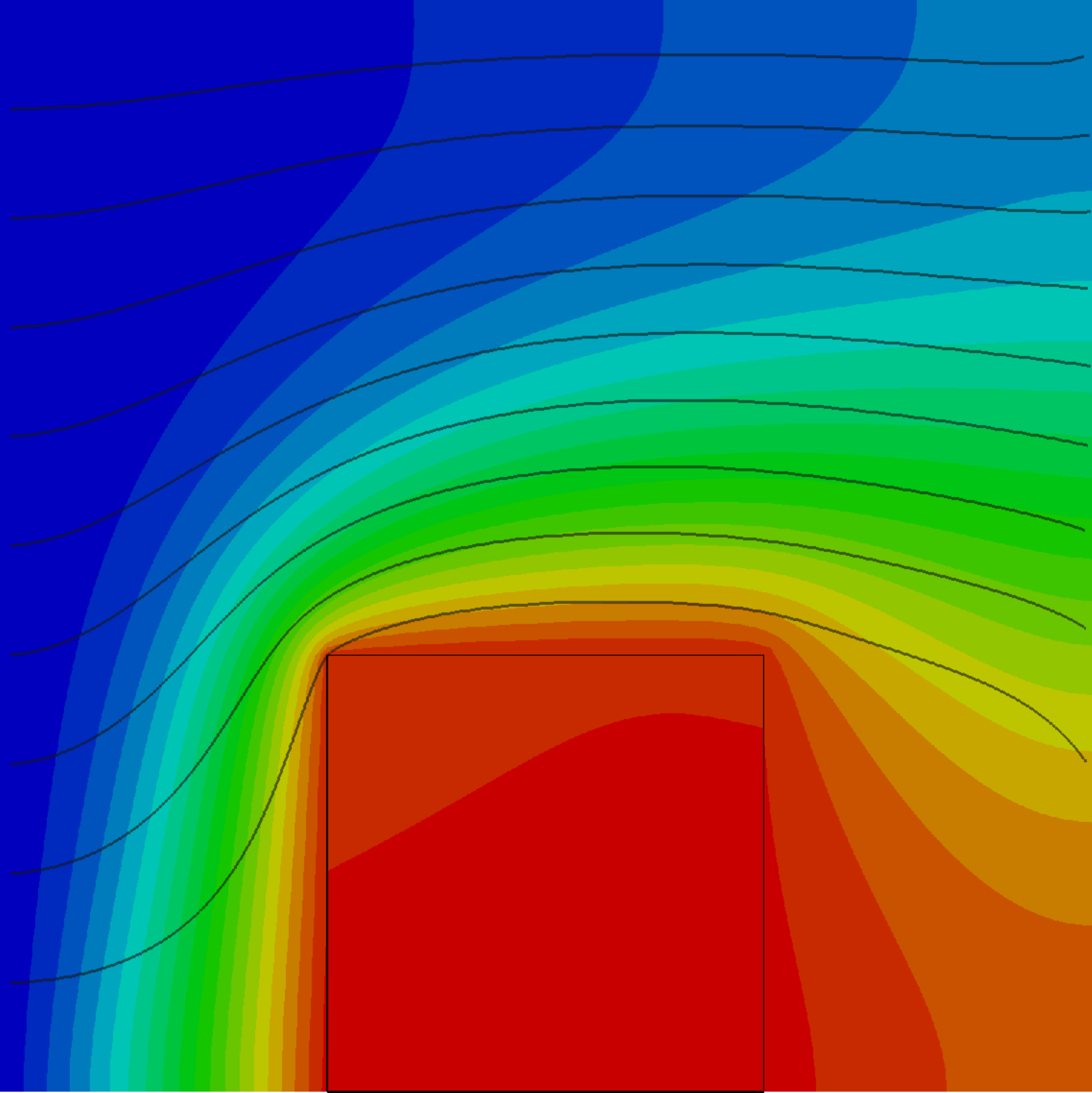}
\label{fig:convection_illustration-a}}
\hspace*{0.03\textwidth}
\subfloat[Natural convection]{\includegraphics[height=0.3\textwidth]{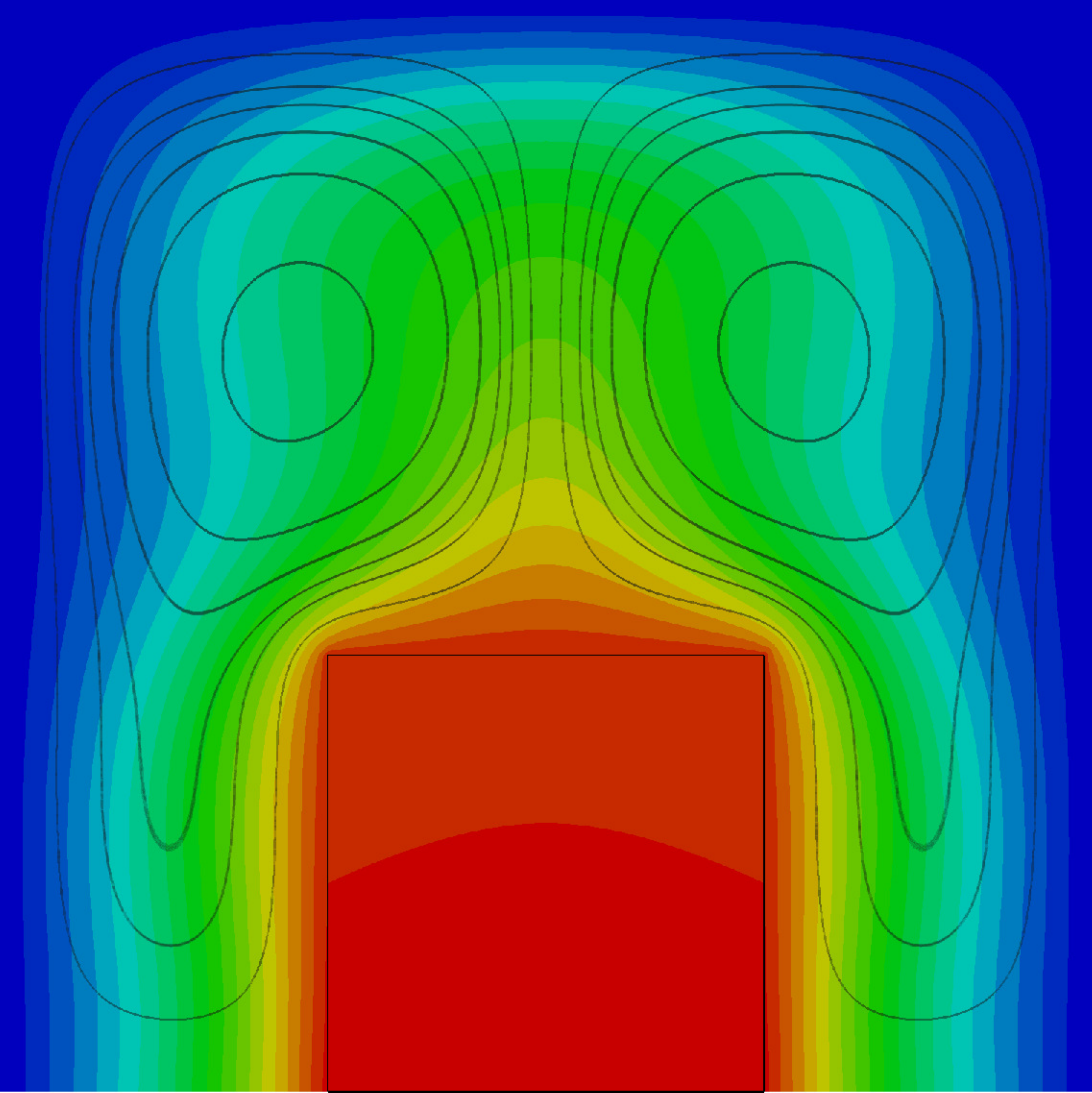}
\label{fig:convection_illustration-b}}
\hspace*{0.03\textwidth}
\subfloat[Diffusion]{\includegraphics[height=0.3\textwidth]{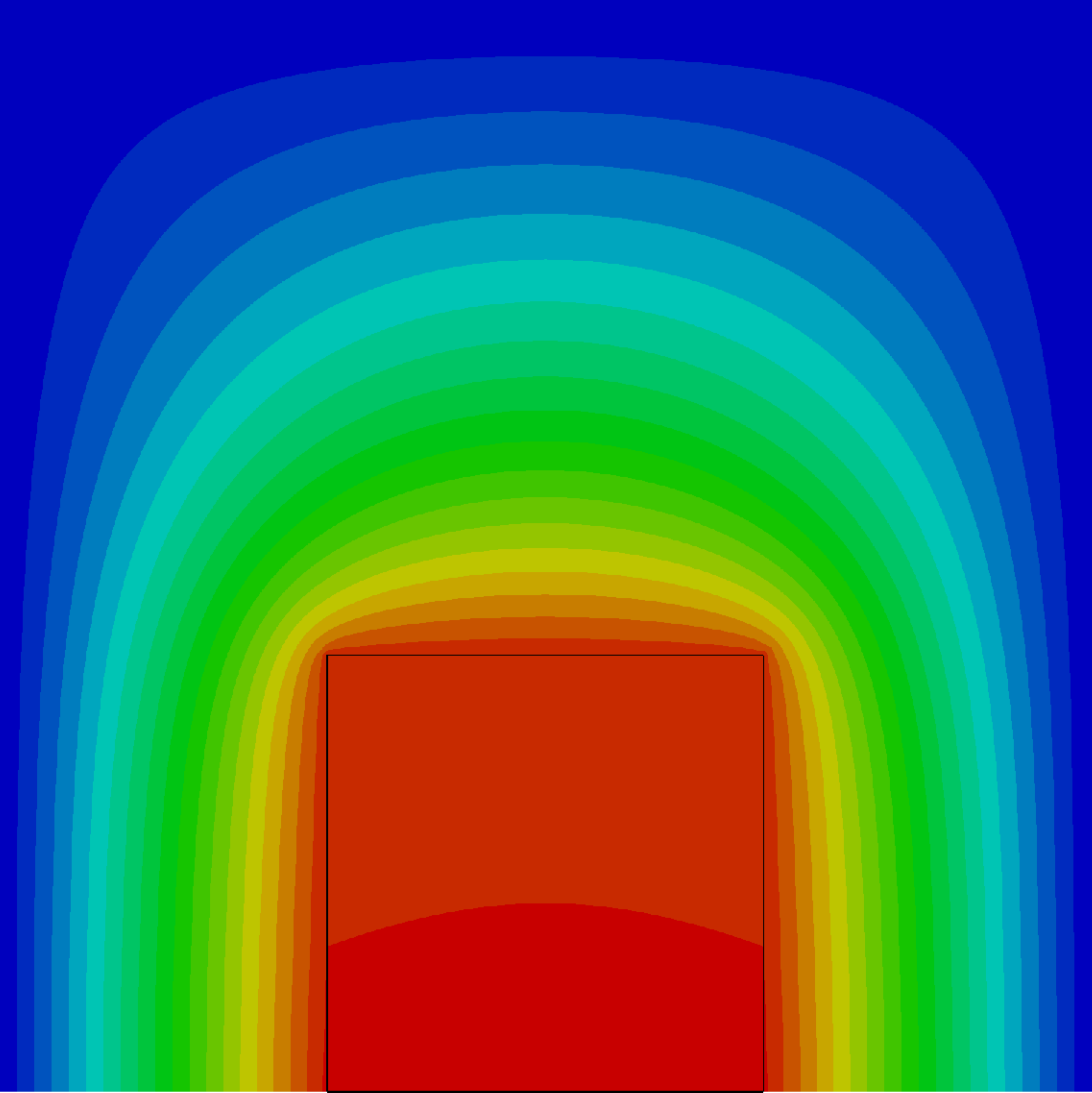}
\label{fig:convection_illustration-c}}
\caption{Illustration of a metallic block subjected to different heat transfer mechanisms in the surrounding fluid. Subfigure {(a)} shows forced convection with a cold flow entering at the left-hand side. Subfigures {(b)} and {(c)} show natural convection and pure diffusion, respectively, due to cold upper and side walls. } \label{fig:convection_illustration}
\end{figure}

The above references on convective heat transfer problems are all concerned with forced convection, where the fluid motion is driven by e.g. a fan, pump or pressure-gradient, as illustrated in figure \ref{fig:convection_illustration-a}. This work applies topology optimisation to natural convection problems, where the fluid motion is governed by differences in buoyancy arising from temperature gradients. This means that the state problem is fully coupled, where the temperature field gives rise to fluid motion as well as the fluid motion affecting the temperature field through convection, as can be seen in figure \ref{fig:convection_illustration-b}.
To the authors' knowledge this has not been done before in the published literature. 
The developed methodology is applied to both a heat sink problem, where the solid temperature field is in focus, and a micropump problem, where the fluid velocity is in focus.

The paper is organised as follows: Section \ref{sec:govequ} describes the governing equations and assumptions, section \ref{sec:FEM} describes the stabilised finite element formulation, section \ref{sec:topopt} introduces the topology optimisation problem and methodology and section \ref{sec:implem} covers the details of the implementation. Finally, section \ref{sec:examp} presents numerical design examples and section \ref{sec:discconc} contains a discussion and conclusions.

\vspace{-6pt}

\section{Governing equations} \label{sec:govequ}
\vspace{-2pt}

\begin{figure}
\centering
\includegraphics[width=0.45\textwidth]{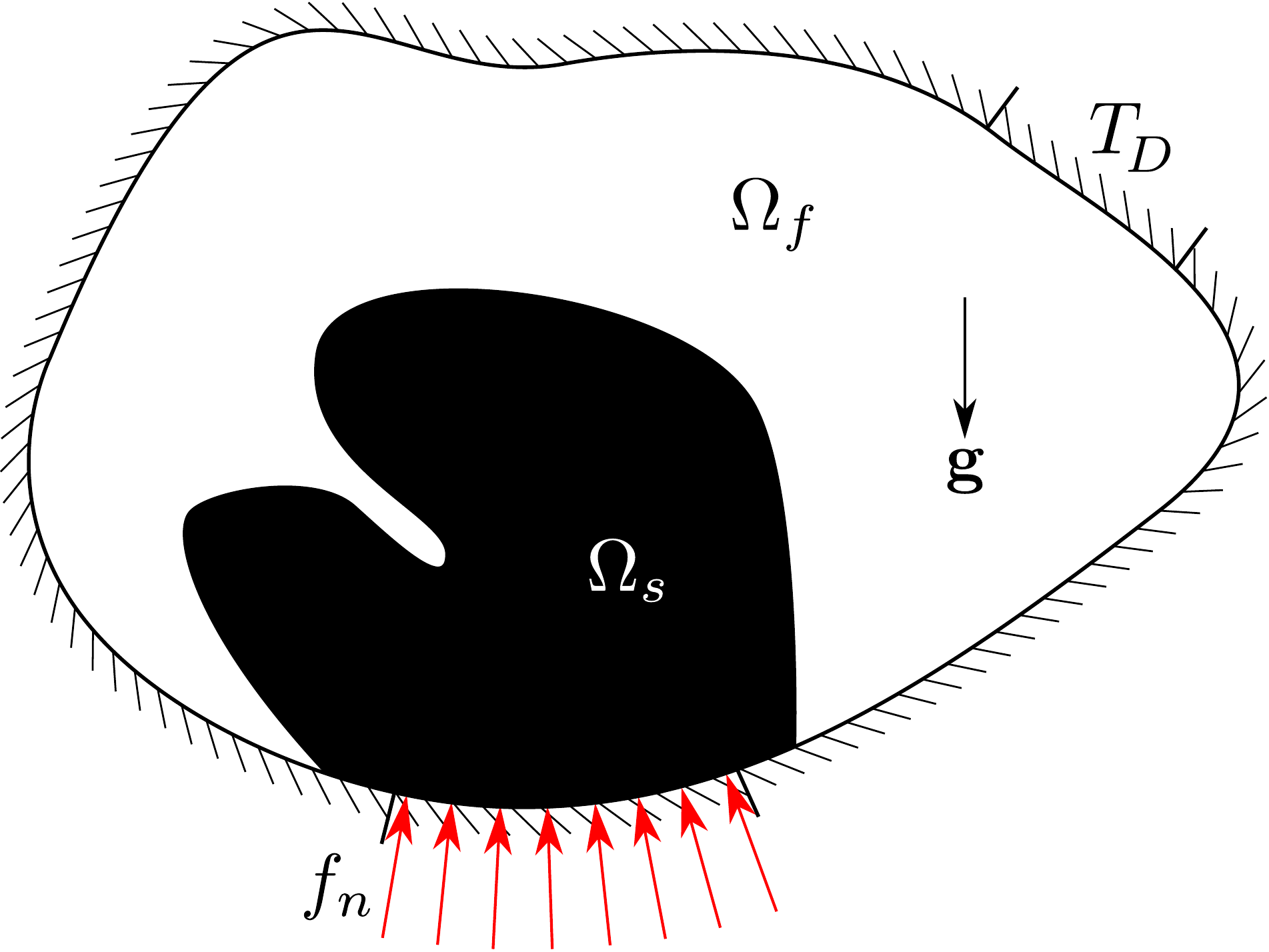}
\caption{Arbitrary enclosed domain consisting of a solid domain, $\Omega_{s}$, engulfed by a fluid domain, $\Omega_{f}$, subjected to a specified temperature, $T_{D}$, a specified heat flux, $f_{n}$, and a gravitational vector $\vecr{g}$.} \label{fig:arbitrary}
\end{figure}
Figure \ref{fig:arbitrary} shows an arbitrary domain consisting of a solid domain, $\Omega_{s}$, engulfed by a fluid domain, $\Omega_{f}$. The total domain is enclosed and thus a no-slip condition, $u_{i} = 0$, exists at all external boundaries, where $u_{i}$ is the fluid velocity field. The possible thermal boundary conditions are also shown, which include a specified temperature, $T_{D}$, and a specified heat flux, $f_{n}$. The acceleration due to gravity is characterised by the vector $\vecr{g}$.

Throughout this paper, the flows are assumed to be steady and laminar. The fluid is assumed to be incompressible, but bouyancy effects are taken into account through the Boussinesq approximation, which introduces variations in the fluid density due to temperature gradients. In order to facilitate the topology optimisation of fluid flow, a Brinkman friction term is introduced.

\subsection{Incompressible and isothermal flow}

Under the assumption of constant fluid properties, incompressible, isothermal and steady flow and neglecting viscous dissipation, the set of equations governing the conservation of momentum, mass and energy for incompressible isothermal steady-state fluid flow are:
\begin{align}
\rho u_{j} \frac{\partial u_{i}}{\partial x_{j}} - \frac{\partial \sigma_{ij}}{\partial x_{j}} &= s_{i} + \rho g_{i} &\,\text{ in }\,\Omega_{f} \label{eq:ConvCollSteadyIncomp-a}\\
\frac{\partial u_{j}}{\partial x_{j}} &= 0 &\,\text{ in }\,\Omega_{f} \label{eq:ConvCollSteadyIncomp-b}\\
\rho c_{p} u_{j} \frac{\partial T}{\partial x_{j}} - k_{f}\frac{\partial^{2} T}{\partial x_{j} \partial x_{j}} &= s_{T} &\,\text{ in }\,\Omega_{f} \label{eq:ConvCollSteadyIncomp-c}
\end{align}
where $\rho$ is the fluid density, $c_{p}$ is the fluid specific heat capacity under constant pressure, $k_{f}$ is the fluid thermal conductivity, $T$ is the temperature field, $x_{i}$ is the spatial coordinate, $s_{i}$ is a momentum source term, $s_{T}$ is a volumetric heat source term, and $\sigma_{ij}$ is the simplified fluid stress tensor given by:
\begin{equation} \label{eq:stress_tensor}
\sigma_{ij} = -\delta_{ij} p + \mu \left( \frac{\partial u_{i}}{\partial x_{j}} + \frac{\partial u_{j}}{\partial x_{i}} \right)
\end{equation}
where $p$ is the pressure, $\mu$ is the fluid dynamic viscosity, and $\delta_{ij}$ is the Kronecker delta.

Equation \eqref{eq:ConvCollSteadyIncomp-a} describes the convection and diffusion of momentum, equation \eqref{eq:ConvCollSteadyIncomp-b} enforces an incompressible and divergence free flow and equation \eqref{eq:ConvCollSteadyIncomp-c} describes the convection and diffusion of thermal energy quantified by the temperature.

\subsection{The Boussinesq approximation}

In order to include buoyancy effects due to temperature differences in the fluid, the Boussinesq approximation will be introduced. It is assumed that the density variations are assumed to be small enough so that they are only important in the volumetric gravity force, which can be written as:
\begin{equation} \label{eq:bouss_def1}
\rho g_{i} = \rho_{0} g_{i} + (\rho - \rho_{0})g_{i}
\end{equation}
where $\rho_{0}$ is a reference fluid density. By assuming only small temperature differences, the density is approximated as a linear function of the temperature around the reference fluid density. This yields the Boussinesq approximation:
\begin{equation} \label{eq:bouss_definiton}
\rho g_{i} \approx \rho_{0} g_{i} \left(1 - \beta\, \left(T-T_{0}\right) \right)
\end{equation}
where $\beta$ is the coefficient of thermal volume expansion and $T_{0}$ is the temperature corresponding to the reference density.
Inserting equation \eqref{eq:bouss_definiton} into equation \eqref{eq:ConvCollSteadyIncomp-a} yields the set of equations governing the conservation of momentum, mass and energy for incompressible steady-state fluid flow taking buoyancy effects into account:
\begin{align}
\rho_{0} u_{j} \frac{\partial u_{i}}{\partial x_{j}} - \frac{\partial \sigma_{ij}}{\partial x_{j}} &= s_{i} - \rho_{0}\,g_{i}\,\beta\left(T-T_{0}\right) &\,\text{ in }\,\Omega_{f} \label{eq:BoussCollSteadyIncomp-a}\\
\frac{\partial u_{j}}{\partial x_{j}} &= 0 &\,\text{ in }\,\Omega_{f} \label{eq:BoussCollSteadyIncomp-b}\\
\rho_{0} c_{p} u_{j} \frac{\partial T}{\partial x_{j}} - k_{f}\frac{\partial^{2} T}{\partial x_{j} \partial x_{j}} &= s_{T} &\,\text{ in }\,\Omega_{f} \label{eq:BoussCollSteadyIncomp-c}
\end{align}
where the constant gravitational body force, in equation \eqref{eq:bouss_definiton}, has been absorbed into the pressure, by using the fact that gravity is a conservative force and it therefore can be represented as the gradient of a scalar quantity, modifying the pressure to include the so-called ``gravitational head''.

\subsection{Dimensionless form}\label{sec:govequ_dimless}

The dimensionless form of the governing equations are used as a basis for the finite element formulation described in section \ref{sec:FEM}. The following relations are used to non-dimensionalise the governing equations:
\begin{align}
u_{i} &= U u_{i}^{*} \label{eq:nondims_u}\\
x_{i} &= L x_{i}^{*} \label{eq:nondims_x}\\
p &= \rho_{0} U^{2} p^{*} \label{eq:nondims_p}\\
s_{i} &= \rho_{0} \frac{U^{2}}{L} s_{i}^{*} \label{eq:nondims_si}\\
T &= \Delta T\,T^{*} + T_{0} \label{eq:nondims_T}\\
s_{T} &= \rho_{0} c_{p} \Delta T \frac{U}{L} s_{T}^{*} \label{eq:nondims_sT}
\end{align}
where \textit{U} is a reference velocity, \textit{L} is a reference length, $\Delta T$ is a reference temperature difference and all variables marked with an asterisks are dimensionless quantities. As this paper deals with pure natural convection problems, where there is no forcing velocity present, the reference velocity $U$ is defined as the diffusion velocity:
\begin{equation} \label{eq:difvel}
U = \frac{\Gamma}{L}
\end{equation}
where $\Gamma$ is the thermal diffusivity of the fluid, defined as:
\begin{equation}
\Gamma = \frac{k_{f}}{\rho_{0} c_{p}}
\end{equation}

By combining equations \eqref{eq:stress_tensor}, (\ref{eq:BoussCollSteadyIncomp-a}-\ref{eq:BoussCollSteadyIncomp-c}) and (\ref{eq:nondims_u}-\ref{eq:nondims_sT}) and collecting the coefficients, the final dimensionless governing equations become:
\begin{align}
u_{j} \frac{\partial u_{i}}{\partial x_{j}} - {Pr}\frac{\partial}{\partial x_{j}} \left( \frac{\partial u_{i}}{\partial x_{j}} + \frac{\partial u_{j}}{\partial x_{i}} \right) + \frac{\partial p}{\partial x_{i}} &= s_{i} - {Gr}{Pr}^{2}\,e^{g}_{i}\,T &\,\text{ in }\,\Omega_{f} \label{eq:BoussDimless-a}\\
\frac{\partial u_{j}}{\partial x_{j}} &= 0 &\,\text{ in }\,\Omega_{f} \label{eq:BoussDimless-b}\\
u_{j} \frac{\partial T}{\partial x_{j}} - \frac{\partial^{2} T}{\partial x_{j} \partial x_{j}} &= s_{T} &\,\text{ in }\,\Omega_{f} \label{eq:BoussDimless-c}
\end{align}
where $Pr$ is the Prandtl number, $Gr$ is the Grashof number, $e^{g}_{i}$ is the unit vector in the gravitational direction and the asterikses denoting dimensionless quantities have been dropped for convenience.

The Prandtl number is defined as:
\begin{equation}
Pr = \frac{\nu}{\Gamma}
\end{equation}
where $\nu = \frac{\mu}{\rho_{0}}$ is the kinematic viscosity, or momentum diffusivity. The Prandtl number is defined from fluid material constants and describes the ratio between the momentum and thermal diffusivities of the fluid and, thus, the relative spreading of viscous and thermal effects. For $Pr$ below unity, diffusion is more effective for heat transfer than momentum transfer and vice versa for $Pr$ above unity. Liquid metals have small $Pr$, gases slightly less than unity, light liquids somewhat higher than unity and oils very large $Pr$. The same flow can thus exhibit vastly different heat transfer characteristics for fluids of different Prandtl numbers.

The Grashof number is defined as:
\begin{equation}
Gr = \frac{g \beta \,{\Delta T}\,L^{3}}{\nu^{2}}
\end{equation}
and thus describes the ratio between the buoyancy and viscous forces in the fluid. The Grashof number is therefore used to describe to what extent the flow is dominated by natural convection or diffusion. For low $Gr$ the flow is dominated by viscous diffusion and for high $Gr$ the flow is dominated by natural convection.

There exists two critical limits when it comes to natural convection. The first corresponds to the transition from a static state, where the fluid remains still because the buoyancy forces are not large enough to trigger fluid motion, to a state with fluid motion. The second corresponds to the transition from laminar to turbulent natural convection. The problems in this paper are assumed to be in the interval between these two critical points and thus exhibit laminar fluid motion. There has been some investigation into these critical limits, for instance for a cylinder heated from below where the critical Grashof numbers are $Gr_{c,1}{Pr} \approx 5\cdot 10^{3} $ and  $Gr_{c,2}{Pr} \approx 10^{5}$ for an aspect ratio of 1 \cite{Priye2012}. However, these critical limits are highly dependent on the geometry and boundary conditions and can thus not be used to validate the assumptions for the complex and changing geometries that can arise during topology optimisation. This would require experimental investigations for the specific problem and that is beyond the scope of this paper.

\subsection{Brinkman friction term}

The Brinkman friction term is introduced in order to facilitate the topology optimisation of fluid flow problems \cite{Borrvall2003}. The Brinkman friction term is a velocity-dependent momentum-sink term and represents the friction force exerted on a fluid flow when passing through an idealised porous medium. The Brinkman friction term is defined as:
\begin{equation}
s_{i} = -\alpha u_{i}
\end{equation}
where $\alpha$ is the effective inverse permeability of the porous medium. It is defined as:
\begin{equation}
\alpha = \frac{Pr}{Da}
\end{equation}
where $Da$ is the Darcy number.
The Darcy number is defined as the dimensionless permeability of the porous medium:
\begin{equation}
Da = \frac{\kappa}{L^{2}}
\end{equation}
where $\kappa$ is the dimensional permeability of the porous medium and $L$ is the reference lengthscale from the non-dimensionalisation process, equation \eqref{eq:nondims_x}.

In order to effectively simulate an immersed solid body inside of a fluid flow, one would ideally set $\alpha = \infty$, or $Da = 0$, inside the solid domain in order not to allow fluid to pass through it at all. However, this is not possible numerically and a large value, $\alpha \gg 1$, is therefore used instead. This value has to be chosen carefully and must be sufficiently large to ensure negligibly small velocities in the solid domain, while small enough so as to ensure numerical stability and stable optimisation convergence.
Throughout this paper, $\alpha$ is set to zero in the fluid parts of the domain, recovering the original Navier-Stokes equations with Boussinesq approximation. This is equivalent to assuming an infinite domain in the out-of-plane direction for the two-dimensional problems considered.

\subsection{Thermal conduction in a solid}\label{sec:govequ_conduc}

The heat transfer through a solid is governed by the diffusion, or conduction, process. The steady-state governing equation for the heat transfer within a solid, the heat conduction equation, is thus very similar to the fluid equivalent, except for the lack of a convective term:
\begin{equation}
- k_{s}\frac{\partial^{2} T}{\partial x_{j} \partial x_{j}} = s_{T} \,\text{ in }\,\Omega_{s} \label{eq:ConducSteady}
\end{equation}
where $k_{s}$ is the thermal conductivity of the solid material. By non-dimensionalising the conduction equation using the relations in equations \eqref{eq:nondims_x}, (\ref{eq:nondims_T}-\ref{eq:nondims_sT}) and \eqref{eq:difvel}, the following dimensionless equation is obtained:
\begin{equation}
- \frac{1}{C_{k}}\frac{\partial^{2} T}{\partial x_{j} \partial x_{j}} = s_{T} \,\text{ in }\,\Omega_{s} \label{eq:ConducDimless}
\end{equation}
where:
\begin{equation}
C_{k} = \frac{k_{f}}{k_{s}}
\end{equation}
is the ratio between the thermal conductivities of the fluid and solid materials.

Equations \eqref{eq:BoussDimless-c} and \eqref{eq:ConducDimless} can thus be collected to a single unifying equation:
\begin{equation}
u_{j} \frac{\partial T}{\partial x_{j}} - K\negthinspace(\vecr{x}) \frac{\partial^{2} T}{\partial x_{j} \partial x_{j}} = s_{T} \,\text{ in }\,\Omega_{f} \cup \Omega_{s}
\end{equation}
where it is assumed that the velocities $u_{j}$ are zero in $\Omega_{s}$ and $K(\vecr{x})$ is the effective conductivity given by:
\begin{equation}
K\negthinspace(\vecr{x}) = \left\lbrace
\begin{matrix}
1 & \text{ if }\, \vecr{x} \in \Omega_{f} \\
\frac{1}{C_{k}} & \text{ if }\, \vecr{x} \in \Omega_{s}
\end{matrix} \right.
\end{equation}
In practice, the assumption of zero velocities is approximately fulfilled by penalising the velocities inside the solid domain, $\Omega_{s}$, using the Brinkman friction term.

\vspace{-6pt}

\section{Finite element formulation}  \label{sec:FEM}
\vspace{-2pt}

The governing equations are discretised using stabilised bilinear quadrilateral finite elements. The standard Galerkin finite element method runs into problems when used with certain unstable combinations of finite element spaces for velocity and pressure, as well as for convection-dominated problems. The addition of stabilisation terms to the weak form equations are therefore neccessary to ensure smooth non-oscillatory solutions.

The Pressure Stabilising Petrov-Galerkin (PSPG) method is used in order to allow for the use of equal-order interpolation for the velocity and pressure fields. The PSPG stabilisation was first introduced for the Stokes equations by Hughes et al. \cite{Hughes1986a} and later generalised for the incompressible Navier-Stokes equations by Tezduyar et al. \cite{Tezduyar1992a} and has since seen widespread use in the finite element modelling of fluid flow. The PSPG stabilisation affects the discrete continuity equation and allows otherwise unstable elements to circumvent the Ladyzhenskaya-Babu{s}ka-Brezzi (LBB), or inf-sup, stability condition for the finite element spaces \cite{Hughes1986a}. This condition is satisfied by a range of different combinations of finite element spaces, such as second-order and first-order interpolation for the velocity and pressure fields, respectively. But using higher-order elements for topology optimisation quickly becomes computationally expensive, as here one ideally wants to refine the mesh quite heavily in order to capture the design with a high resolution.

Furthermore, the Streamline-Upwind Petrov-Galerkin (SUPG) method,  as presented by Brooks and Hughes \cite{Brooks1982}, is used in order to supress oscillations in the velocity and temperature fields due to sharp solution gradients in the streamline direction, which often arise in convection-dominated problems due to downstream boundary conditions. The SUPG stabilisation method can be seen as a generalisation of upwinding schemes in finite difference and finite volume methods, adding a carefully scaled amount of numerical diffusion in the streamline direction.

To obtain the finite element discretised equations, the weak form of the governing equations is found by multiplying the strong form, equations (\ref{eq:BoussDimless-a}-\ref{eq:BoussDimless-c}), with suitable test functions and integrating over the domain. The suitable finite dimensional spaces, $\vecrsym{\mathcal{U}}^{h}$, $\mathcal{S}^{h}$, $\vecrsym{\mathcal{W}}^{h}$, $\mathcal{V}^{h}$ and $\mathcal{Q}^{h}$, are introduced and the discrete variational problem becomes:
\newline \newline
\noindent Find $\vecr{u}^{h} \in \vecrsym{\mathcal{U}}^{h}$, $p^{h} \in \mathcal{Q}^{h}$ and $T^{h} \in \mathcal{S}^{h}$ such that $\forall \, \vecr{w}^{h} \in \vecrsym{\mathcal{W}}^{h} $, $\forall \, q^{h} \in \mathcal{Q}^{h}$ and $\forall \, v^{h} \in \mathcal{V}^{h} $:
\vspace*{-0.1cm}
\begin{equation}\label{eq:weakform_mom}
\begin{split}
\overbrace{\int_{\Omega} w_{i}^{h} u_{j}^{h} \frac{\partial u_{i}^{h} }{\partial x_{j}} \, dV }^{\text{convection}} + \overbrace{ \int_{\Omega} \frac{\partial w_{i}^{h} }{\partial x_{j}} \,{Pr} \left( \frac{\partial u_{i}^{h} }{\partial x_{j}} + \frac{\partial u_{j}^{h} }{\partial x_{i}} \right)\, dV }^{\text{diffusion}} - \overbrace{ \int_{\Omega} w_{i}^{h} \frac{\partial p^{h}}{\partial x_{i}} \, dV}^{\text{pressure coupling}} + \overbrace{ \int_{\Omega} w_{i}^{h} \alpha u_{i}^{h}\, dV }^{\text{Brinkman friction}} \\ + \underbrace{ \int_{\Omega} w_{i}^{h} \,{Gr}\,{Pr}^{2}\,e_{i}^{g} T\, dV }_{\text{Boussinesq force}} - \underbrace{ \int_{\Gamma^{u}_{N}} w_{i}^{h} h_{i}\, dS }_{\text{surface traction}} + \underbrace{ \sum_{e=1}^{N_{e}} \int_{\Omega_{e}} \tau_{SU} u_{j}^{h} \frac{\partial w_{i}^{h}}{\partial x_{j}} R^{u}_{i}(\mathbf{u}^{h},p^{h},T^{h}) \, dV}_{\text{SUPG stabilisation}} = 0
\end{split}
\end{equation}
\begin{equation}\label{eq:weakform_cont}
- \underbrace{ \int_{\Omega} q^{h}\, \frac{\partial u_{i}^{h}}{\partial x_{i}} \,dV }_{\text{continuity}} + \underbrace{ \sum_{e=1}^{N_{e}} \int_{\Omega_{e}} \tau_{PS} \frac{\partial q^{h}}{\partial x_{i}} R^{u}_{i}(\mathbf{u}^{h},p^{h},T^{h}) \, dV}_{\text{PSPG stabilisation}} = 0
\end{equation}
\begin{equation}\label{eq:weakform_condif}
\begin{split}
\underbrace{ \int_{\Omega} v^{h} u_{j}^{h} \frac{\partial T^{h} }{\partial x_{j}} \, dV }_{\text{convection}} &+ \underbrace{ \int_{\Omega} \frac{\partial v^{h} }{\partial x_{j}} K \frac{\partial T^{h} }{\partial x_{j}}\, dV }_{\text{diffusion}} - \underbrace{ \int_{\Omega} v^{h} s_{T}^{h}\, dV }_{\text{volumetric flux}} \\
- \underbrace{ \int_{\Gamma^{T}_{N}} v^{h} f_{n} \, dS }_{\text{surface flux}} &+ \underbrace{ \sum_{e=1}^{N_{e}} \int_{\Omega_{e}} \tau_{SU_{T}} u_{j}^{h} \frac{\partial v^{h}}{\partial x_{j}} R_{T}(\mathbf{u}^{h},T^{h}) \, dV}_{\text{SUPG stabilisation}} = 0
\end{split}
\end{equation}
where $h_{i}$ is the surface traction vector on the surface $\Gamma^{u}_{N}$, $f_{n}$ is the surface heat flux normal to the surface $\Gamma^{T}_{N}$, $R^{u}_{i}$ is the residual form of equation \eqref{eq:BoussDimless-a}, $R_{T}$ is the residual form of equation \eqref{eq:BoussDimless-c}, $\tau_{SU}$ is the SUPG stabilisation parameter for the momentum equation, $\tau_{PS}$ is the PSPG stabilisation parameter and $\tau_{SU_{T}}$ is the SUPG stabilisation parameter for the temperature equation. The stabilisation parameters are described in appendix \ref{sec:stabpar}.

In order for the stabilised discrete weak form equations to be mathematically consistent with the original equations, the stabilisation terms are posed as dependent on the strong form residual. This is what makes them Petrov-Galerkin methods, where the test functions are perturbed. This ensures that a solution to the original problem remains a solution to the stabilised equations.

\vspace{-6pt}

\section{Topology optimisation} \label{sec:topopt}
\vspace{-2pt}
Several optimisation problems are considered in this paper, so a general topology optimisation problem is defined as:
\begin{align} \label{eq:topopt_prob}
\underset{ \vecrsym{\gamma} \in \mathbb{R}^{ n_{\! d} } }{\text{minimise:}} & f\negthinspace \left( \vecrsym{{\gamma}}, \vecr{s} \right)  \nonumber\\
\text{subject to: } & g_{i}\negthinspace \left( \vecrsym{{\gamma}}, \vecr{s} \right) \leq 0 \,\, \text{ for } i = 1,...,m  \\
 & \vecrsym{\mathscrbf{R}}\negmedspace \left( \vecrsym{{\gamma}}, \vecr{s} \right) = \vecr{0} \nonumber \\
 & 0 \leq \gamma_{i} \leq 1 \,\,\,\, \text{ for } i = 1,...,n_{d}  \nonumber
\end{align}
where $\vecrsym{\gamma}$ is a vector of the $n_{d}$-number of design variables, $\vecr{s}$ is a vector of the $n_{s}$-number of state field variables, $f$ is the objective functional, $g_{i}$ are the \textit{m}-number of constraint functionals and 
$\vecrsym{\mathscrbf{R}}\negmedspace \left( \vecrsym{{\gamma}}, \vecr{s} \right) = \matx{M}\negthinspace \left( \vecrsym{{\gamma}}, \vecr{s} \right) \vecr{s} - \vecr{b}\negthinspace \left( \vecrsym{{\gamma}}, \vecr{s} \right)$
 is the residual of the discretised system of equations arising from equations (\ref{eq:weakform_mom} - \ref{eq:weakform_condif}). The optimisation problems are solved using the nested formulation, where the discretised system of equations for the state field is solved separately from the design problem.

\subsection{Interpolation}
The goal of topology optimisation is most often to end up with binary designs, that is where the design variables either take the value 0, representing solid, or 1, representing fluid. Thus, it is important that the physical modelling is correct for these two extremes in order for the final optimised design to be physically realistic. However, when performing topology optimisation with continuous variables, the interpolation between the two extremes is also of utmost importance. It can be discussed whether the intermediate regions, where the design variables take values between 0 and 1, should be physically realistic or not. But when the goal is to have binary designs, the most important thing is to make sure that the intermediate regions are unattractive with respect to the optimisation problem. This is usually done by penalising the intermediate densities with respect to the material parameters, such as impermeability and effective conductivity.

In order to minimise the number of physical properties to interpolate, the Boussinesq forcing term is left constant, with respect to design variable, throughout. This has worked very well for the heat sink problem, but difficulties were faced for the micropump problem, as will be discussed in sections \ref{sec:examp} and \ref{sec:discconc}.

The inverse permeability is interpolated using the following function:
\begin{equation} \label{eq:alphaint}
\alpha\left( \gamma \right) = \underline{\alpha} + \left( \overline{\alpha} - \underline{\alpha} \right) \frac{1-\gamma}{1+q_{\alpha}\gamma}
\end{equation}
which is a reformulated version of the original convex interpolation function as laid out by Borrvall and Petersson in their seminal paper \cite{Borrvall2003}. The convexity factor, $q_{\alpha} \geq 0$, determines the convexity of the function and can thus be adjusted to determine the effective inverse permeability for the intermediate design variables.

The difference in the thermal conductivities of the fluid and solid phases is included through interpolation of the effective conductivity, $K$, which was defined in section \ref{sec:govequ_conduc}. The effective conductivity is interpolated using the following function:
\begin{equation} \label{eq:conducint}
K( \gamma ) = \frac{\gamma ( C_{k} ( 1 + q_{f} ) -1 ) + 1}{ C_{k} ( 1 + q_{f} \gamma ) }
\end{equation}
which is a RAMP-style function \cite{Stolpe2001a}. The convexity factor, $q_{f} \geq 0$, can be adjusted in order to penalise intermediate design variables with respect to effective conductivity and, thus, forcing the design variables towards the bounds of 0 and 1.

\subsection{Adjoint sensitivity analysis}

In order to apply gradient-based optimisation algorithms to the topology optimisation problem \eqref{eq:topopt_prob}, the gradients of the objective functional and any given constraint functionals need to be known. These gradients, also known as sensitivities, are here found using the discrete adjoint method, see e.g. \cite{Bendsoee2003,Michaleris1994}. This gives rise to the following adjoint problem:
\begin{equation}\label{eq:AdjDiscProb}
{{\frac{\partial\vecrsym{\mathscrbf{R}}}{\partial\vecr{s}}}^{\text{T}}}{\vecrsym{\lambda}} = {{\frac{\partial\Phi}{\partial\vecr{s}}}^{\text{T}}}
\end{equation}
where $\vecrsym{\lambda}$ is the vector of adjoint variables and $\Phi$ is a generic functional that depends on the state and design variables, $\Phi ( \vecrsym{\gamma}, \vecr{s} )$. The sensitivities can easily be calculated as:
\begin{equation}\label{eq:AdjSensFinal}
{\frac{d\Phi}{d\vecrsym{\gamma}}} = {\frac{\partial\Phi}{\partial\vecrsym{\gamma}}} - {\vecrsym{\lambda}^{\text{T}}} {\frac{\partial\vecrsym{\mathscrbf{R}}}{\partial\vecrsym{\gamma}}}
\end{equation}
where $\frac{d \square }{d \square }$ denotes the total derivative and $\frac{\partial \square }{\partial \square }$ denotes the partial derivative. The partial derivatives of the objective function and residual vector, with respect to the design variables, are derived analytically.

It can be seen that the adjoint problem \eqref{eq:AdjDiscProb} depends on the transpose of the tangent system matrix of the original state problem. For many problems, e.g. linear elasticity and Stokes flow, this matrix is symmetric and the factorisation from solving the state problem can be reused. However, the FEM discretised flow equations result in an unsymmetric tangent system matrix and hence, the transposed matrix must be calculated and factorised before solving the adjoint problem. It is important to note that the adjoint problem is linear, even though the original state equations are non-linear.

It should be noted that sensitivities of all objective and general constraint functionals can be found using the above methodology. A new adjoint variable field is then introduced per functional and thus problem \eqref{eq:AdjDiscProb} needs to be calculated once per functional, where the factorised transposed tangent system matrix can be reused. In this paper, only a simple volume constraint is applied, which does not depend on the state field, and the sensitivities can therefore be derived analytically.

\subsection{Density filter}
Filtering is imposed for the topology optimisation problems with a thermal objective functional. This is done to solve issues with ill-posedness of the optimisation problem and also issues with lengthscale. Fluid flow problems where the objective is to minimise the dissipated energy are generally well-posed and no filtering is needed \cite{Borrvall2003}, but for structural mechanics and heat transfer problems so-called checkerboards may appear in the design solution \cite{Sigmund1998}. 
Filtering is also used to introduce a lengthscale into the design. Mesh-dependency is a well-known issue in topology optimisation for conductive heat transfer and it has also been observed as an issue that thin solid members do not provide enough resistance to effectively inhibit the flow when using the Brinkman approach \cite{Kreissl2012,Alexandersen2013}, so imposing a minimum lengthscale can help on this issue. 

In this work, the density filter \cite{Bruns2001,Bourdin2001} is used, but there exists other methods to fix the complications listed above \cite{Bendsoee2003,Sigmund2007}. The filtered relative densities are defined as a ``weighted average'' of the design variables of the elements within a predefined neighbourhood:
\begin{equation}\label{neigh}
N_{e} = \left\lbrace i \,\lvert\,  \lVert\mathbf{x}_{i}-\mathbf{x}_{e}\lVert \leq R \right\rbrace
\end{equation}
\noindent where \textit{R} is the filter radius and $\mathbf{x}_{i}$ is the spatial location of the element \textit{i}.

It should be noted that the filtered relative densities, $\vecrsym{\tilde{\gamma}}$, become the physically meaningful variables that now replace $\vecrsym{\gamma}$ in the interpolation functions. The now non-physical design variables, $\vecrsym{\gamma}$, are updated using the optimisation algorithm and therefore the sensitivities have to be updated using the chain rule:
\begin{equation}
\frac{\partial \Phi}{\partial\gamma_{e}} =
\sum_{i\in N_{e}} \frac{\partial f}{\partial\tilde{\gamma}_{i}}
\frac{\partial\tilde{\gamma}_{i}}{\partial\gamma_{e}}
\end{equation}

Although filtering solves the problems of checkerboarding and mesh-dependent solutions, it also introduces a band around the edge of the solution where the design transitions from one phase to the other. This can be solved by using projection methods and robust formulations \cite{Sigmund2007,Wang2010} in order to have crisp final designs with a clear separation between material and void. This is left as a subject for future work.

It is important to note that for all the figures showing the design fields, it is the physically relevant filtered relative densities, $\vecrsym{\tilde{\gamma}}$, that are shown.

\vspace{-6pt}

\section{Implementation} \label{sec:implem}
\vspace{-2pt}
The finite element formulation explained in section \ref{sec:FEM} is implemented into DFEM \cite{Aage2013} which is an object-oriented parallel finite element framework programmed in the C++ programming language \cite{Stroustrup}. 

The damped Newton method is used to solve the system of non-linear equations arising from equations (\ref{eq:BoussCollSteadyIncomp-a}-\ref{eq:BoussCollSteadyIncomp-c}) with a constant, experimentally predetermined, damping factor. The size of the damping factor depends on the non-linearity of the system and, thus, the Grashof number. Typically, for low Grashof numbers the full Newton step can be taken, whereas for higher Grashof numbers $20-60\%$ of the step is taken. The reasons why this simple approach is chosen instead of an elaborate update scheme is discussed in section \ref{sec:discconc}. In order to ease the convergence of the non-linear solver for large Grashof numbers, the Grashof number is ramped from a low value and then increased during the non-linear iterations to the required value at intermediate stages of convergence. The multifrontal parallel direct solver MUMPS \cite{Amestoy2000} is used to factorise and solve the linearised system of equations at each non-linear iteration.

For updating the design variables, a parallel implementation \cite{Aage2013} of the Method of Moving Asymptotes \cite{Svanberg1987} is used with a movelimit of $20\%$ and a convergence criteria of $\left| \vecrsym{\gamma}_{k+1} - \vecrsym{\gamma}_{k} \right|_{\infty} < 0.01$, where the subscript denotes design iteration number. In order to ease the convergence of the optimisation, a continuation approach is taken, where the convexity parameter for the effective conductivity, $q_{f}$, is gradually increased during the optimisation process for a constant value of the convexity parameter for the impermeability, $q_{\alpha}$. Unless otherwise stated, the sequence of values are $q_{f} = \left\lbrace 10^{0},10^{1},10^{2},10^{3},10^{4} \right\rbrace$, where the parameter is changed every 100 design iterations or at intermediate stages of convergence and the value of the convexity parameter for the impermeability is $q_{\alpha}=10^{7}$. These values are chosen to aggressively penalise intermediate densities with respect to effective conductivity and to confine the maximum impermeability to the fully solid parts of the domain.

The variation of the stabilisation parameters, described in appendix \ref{sec:stabpar}, due to changes in the state and design fields is ignored when computing the sensitivities of the objective functional. This leads to inconsistent sensitivities, however, finite difference checks show that the adjoint sensitivities are generally very accurate for the thermal compliance functional. For functionals directly dependent on the flow field, slight discrepancies are observed, primarily near boundaries. No oscillatory behaviour has been observed during the optimisation process, so this is not seen as a significant problem for the current examples. However, this can not be guaranteed for all problems, so derivation and implementation of the derivatives of the UGN-based stabilisation parameters is a subject of future research. 

\vspace{-6pt}

\section{Numerical examples} \label{sec:examp}
\vspace{-2pt}

\subsection{Heat sink cooled by natural convection}

\begin{figure}
\centering
\includegraphics[height=0.3\textwidth]{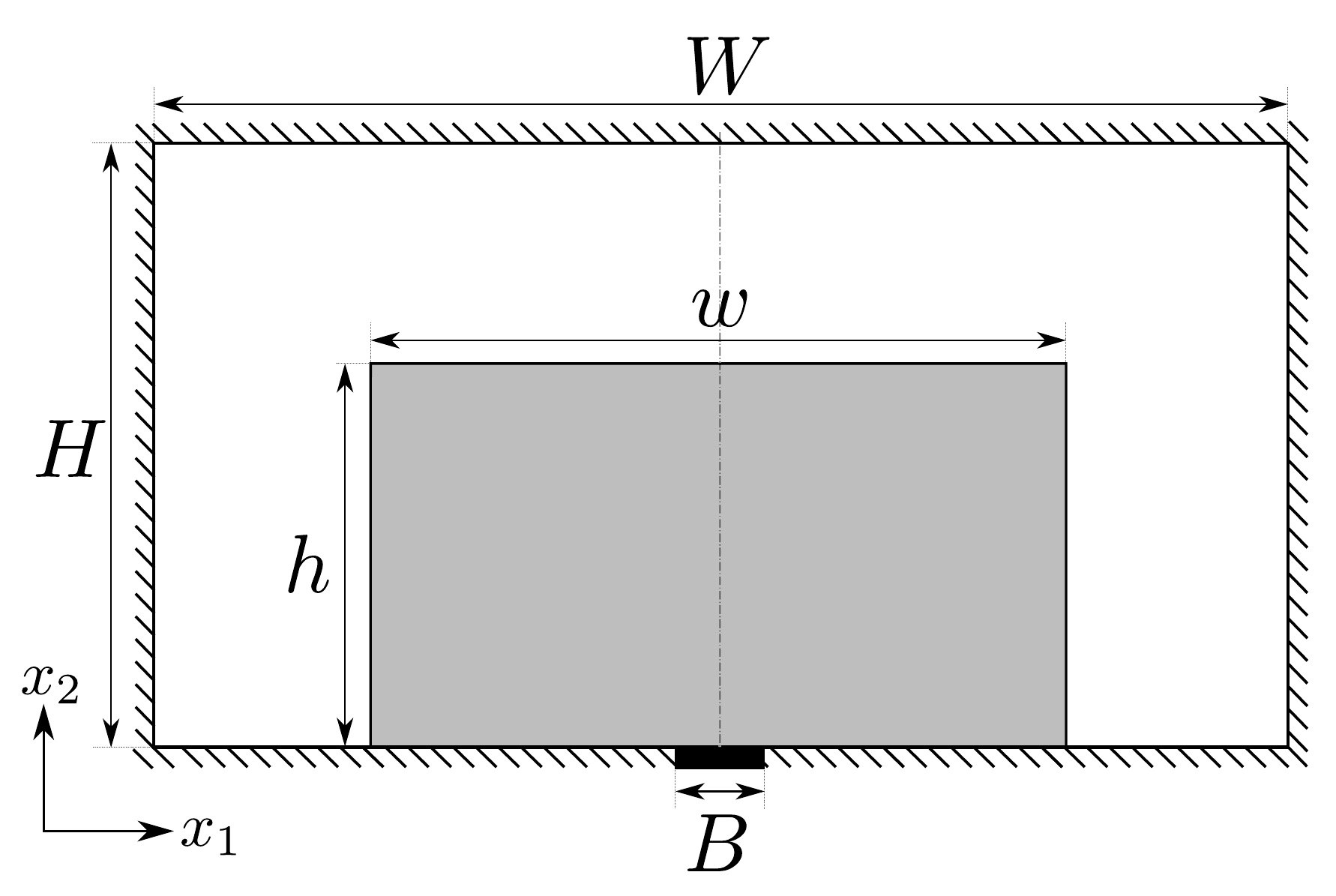}
\includegraphics[height=0.3\textwidth]{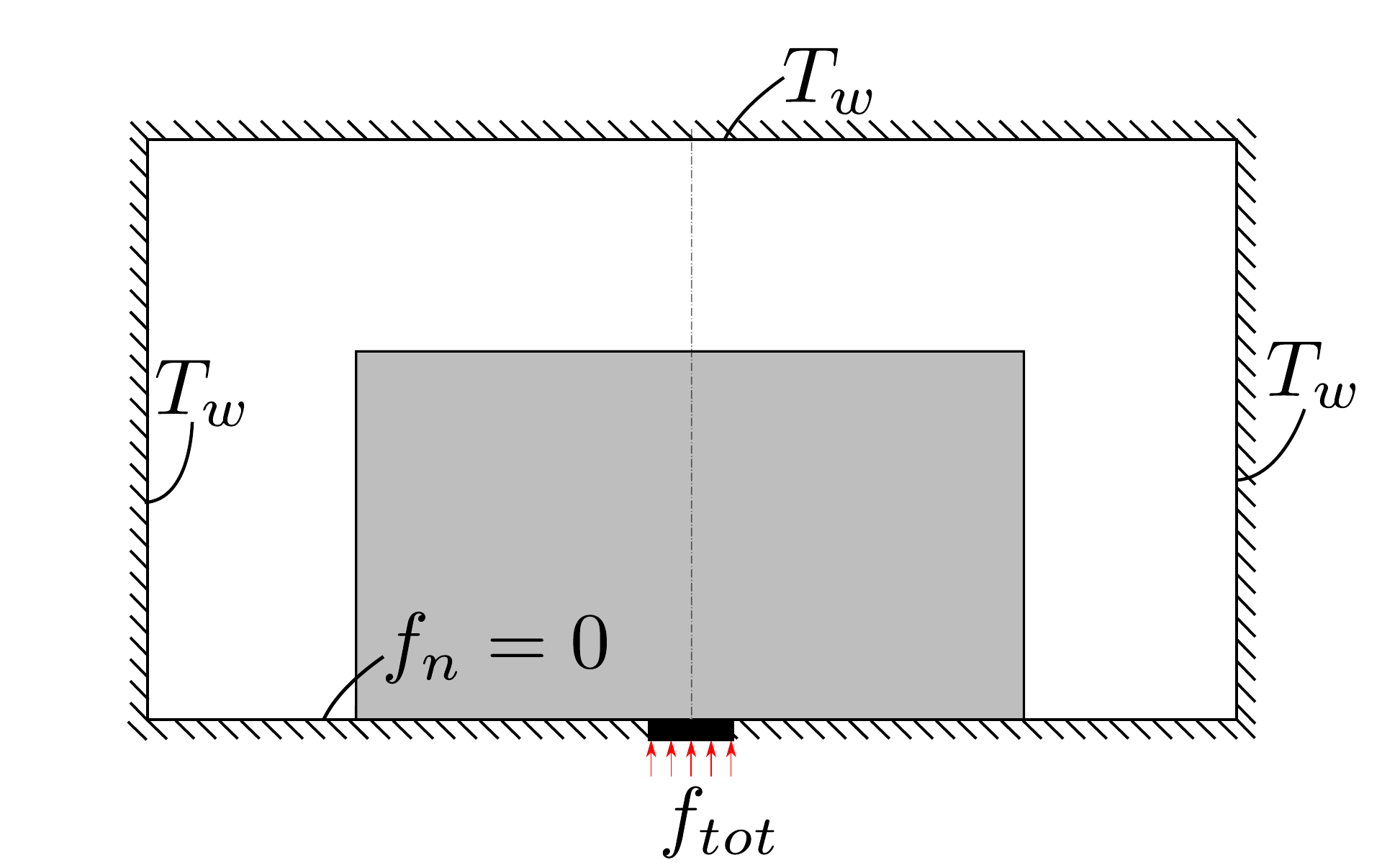}
\includegraphics[height=0.3\textwidth]{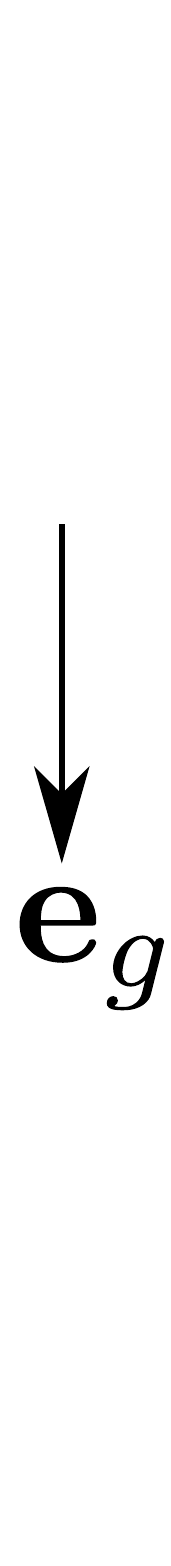}
\caption{Schematic illustration of the layout and boundary conditions for the heat sink subjected to natural convection. Black denotes fully solid, white denotes fully fluid and light grey denotes the design domain.} \label{fig:cpunatconv1_schematic}
\end{figure}
The first numerical example is the design of a heat sink subjected to natural convection cooling due to surrounding cold walls. Figure \ref{fig:cpunatconv1_schematic} shows schematic illustrations of the layout and boundary conditions for the problem. The calculation domain consists of a rectangular design domain on top of a block of solid material that is subjected to a distributed heat flux, $f_{tot}$, along the bottom and a rectangular flow domain surrounding the heat sink. The upper and side walls are kept at a specified temperature, $T_{w}$, and the bottom wall is thermally insulated, $f_{n} = 0$. All walls have no-slip conditions imposed, $u_{i} = 0$.
\begin{table}
\caption{The dimensionless quantities used for the natural convection heat sink problem shown in figure \ref{fig:cpunatconv1_schematic}.}\label{tab:cpunatconv1_quantities}
\centering
\tabsize
\subfloat[Sizes]{
\begin{tabular}{ccccc}
\toprule
Total height & Total width & Design height & Design width & Flux width \\
\midrule
$H=4$ & $W=7$ & $h=2.5$ & $w=4$ & $B=0.2$\\
\bottomrule
\end{tabular} }\\
\tabsize
\subfloat[Boundary conditions]{
\begin{tabular}{cc}
\toprule
Wall temperature & Flux \\
\midrule
$T_{w}=0$ & $f_{tot} = 2.2\cdot10^{-3}$ \\
\bottomrule
\end{tabular} }
\end{table}
Table \ref{tab:cpunatconv1_quantities} lists the dimensionless quantities specifying the layout and boundary conditions of the natural convection heat sink problem.
All of the quantities specified are kept constant throughout. The flow velocities are relative to the diffusion velocity, as explained in section \ref{sec:govequ_dimless}, and the flux and temperature are relative to the scales defined by the non-dimensionalisation process \footnote{The temperature scale for problems with only homogeneous Dirichlet boundary conditions on the temperature, $T=0$, and non-homogeneous Neumann boundary conditions, $f_{n} \neq 0$, is given by $\Delta T = \frac{t L}{k_{s} f_{n}}$ where $t$ is the applied dimensional heat flux.}. The Grashof number, $Gr_{H}$, is based on the height of the entire domain, $H$.

The problem is investigated for varying $Gr_{H}$ under constant parameters, $C_{k}=10^{-2}$, $\overline{\alpha}=10^{7}$ and $Pr=1$. The computational domain, excluding the solid flux base, is discretised using $280\times160$ square elements, where the design domain makes up $160\times100$ of these. The solid flux base is discretised using $8\times4$ elements with only temperature degrees of freedom. The total number of state degrees of freedom for the entire calculation domain is $181,000$. The filter radius is set to 0.06, which is 2.4 times the element size.

The objective functional for the heat sink problem is chosen as the thermal compliance, which has been successfully used as the objective functional in heat transfer problems, e.g. \cite{Bendsoee2003,Yoon2010}. The thermal compliance is defined as:
\begin{equation}
f_{tc} ( \vecrsym{\gamma}, \vecr{t} ) = {\vecr{f}_{t}}^{T} \vecr{t}
\end{equation}
where $\vecr{t}$ is the vector containing the nodal temperature and $\vecr{f}_{t}$ is the heat flux vector from the finite element equations. Thus, by minimising the thermal compliance, the temperatures where heat flux is applied, are minimised and the optimal structure will therefore be one that maximises the transport of thermal energy away from the points of applied heat flux.

When diffusion dominates in the fluid and the conductivity of the solid material is higher than that of the fluid, then the trivial solution is to fill the entire design domain with fully solid material. As the importance of thermal convection increases, it has been observed that the shape of the design becomes more important. However, it has also been observed that as long as the conductivity of the solid material is several orders of magnitude higher than that of the fluid, the optimised design tends to fill the majority of the design domain with solid material for the rather low Grashof numbers considered. Therefore, a constraint on the maximum allowable solid volume fraction is set to $50\%$.

\begin{figure}
\centering
\hspace*{-0.02\textwidth}\subfloat[$Gr_{H} = 640$]{\hspace*{0.095\textwidth}\includegraphics[height=0.25\textwidth]{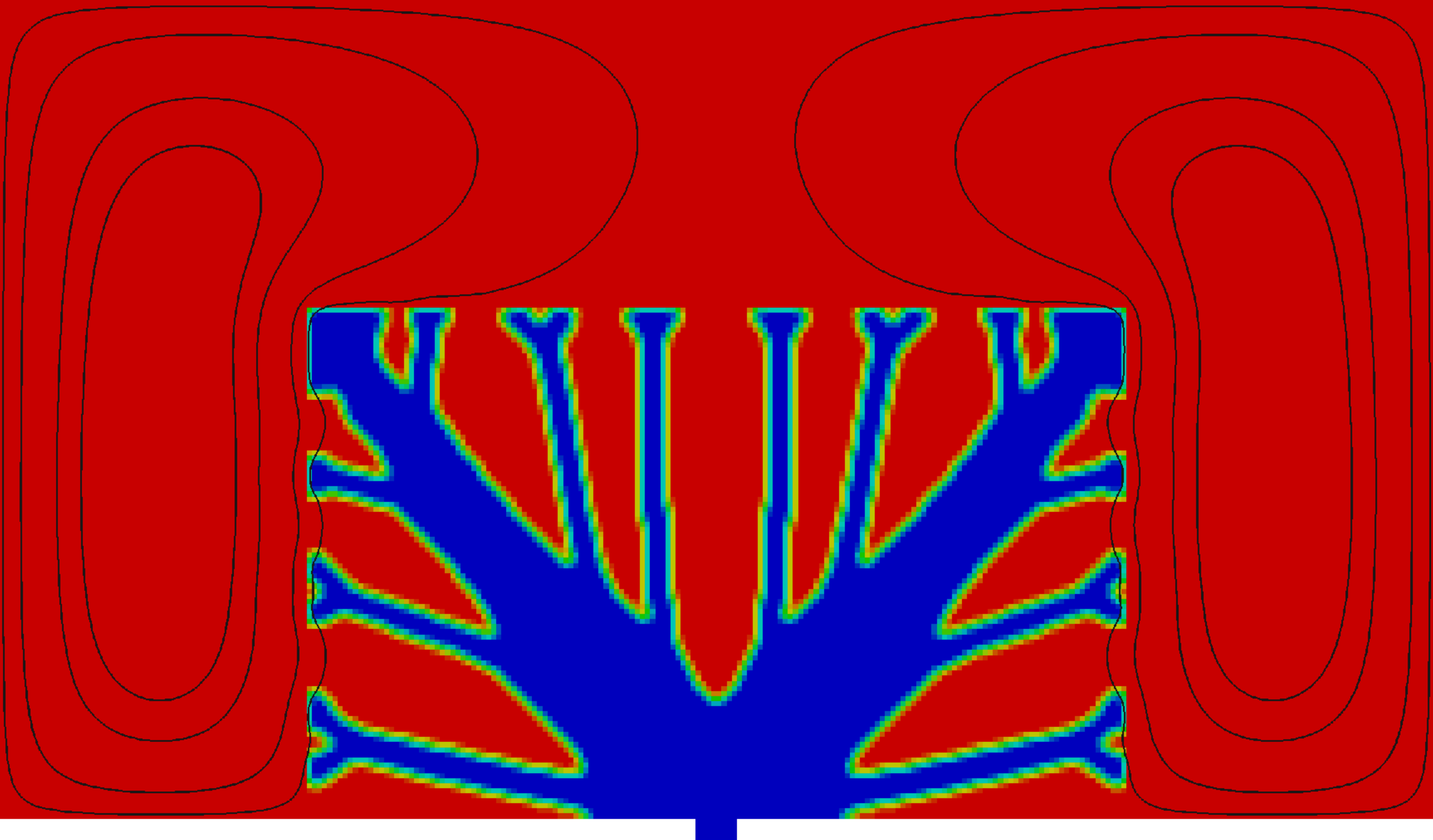}
\hspace*{0.025\textwidth}
\includegraphics[height=0.25\textwidth]{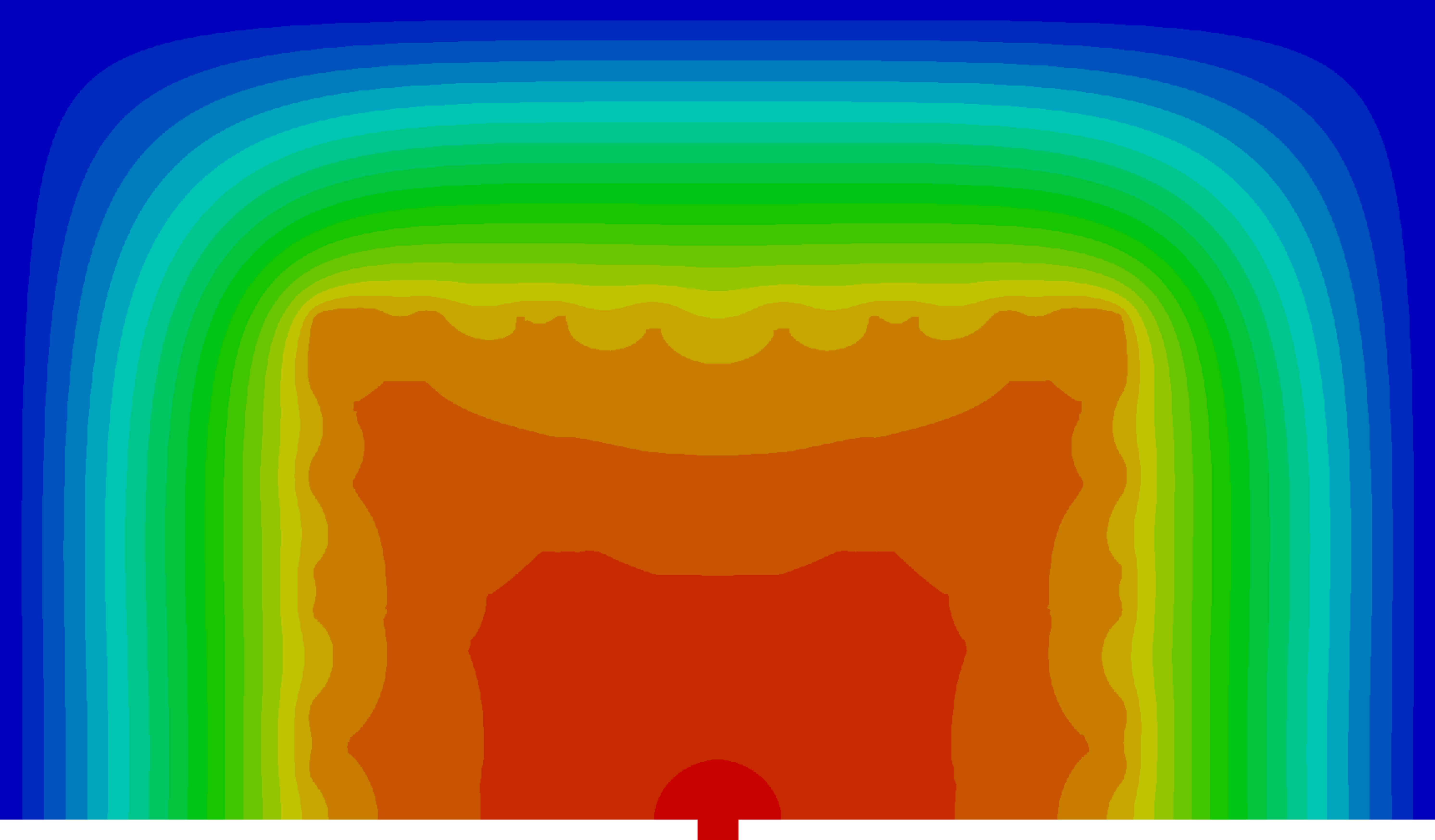}\hspace*{0.095\textwidth} \label{fig:cpunatconv1_streamdes-a}}\\
\hspace*{-0.02\textwidth}\subfloat[$Gr_{H} = 3200$]{\includegraphics[height=0.25\textwidth]{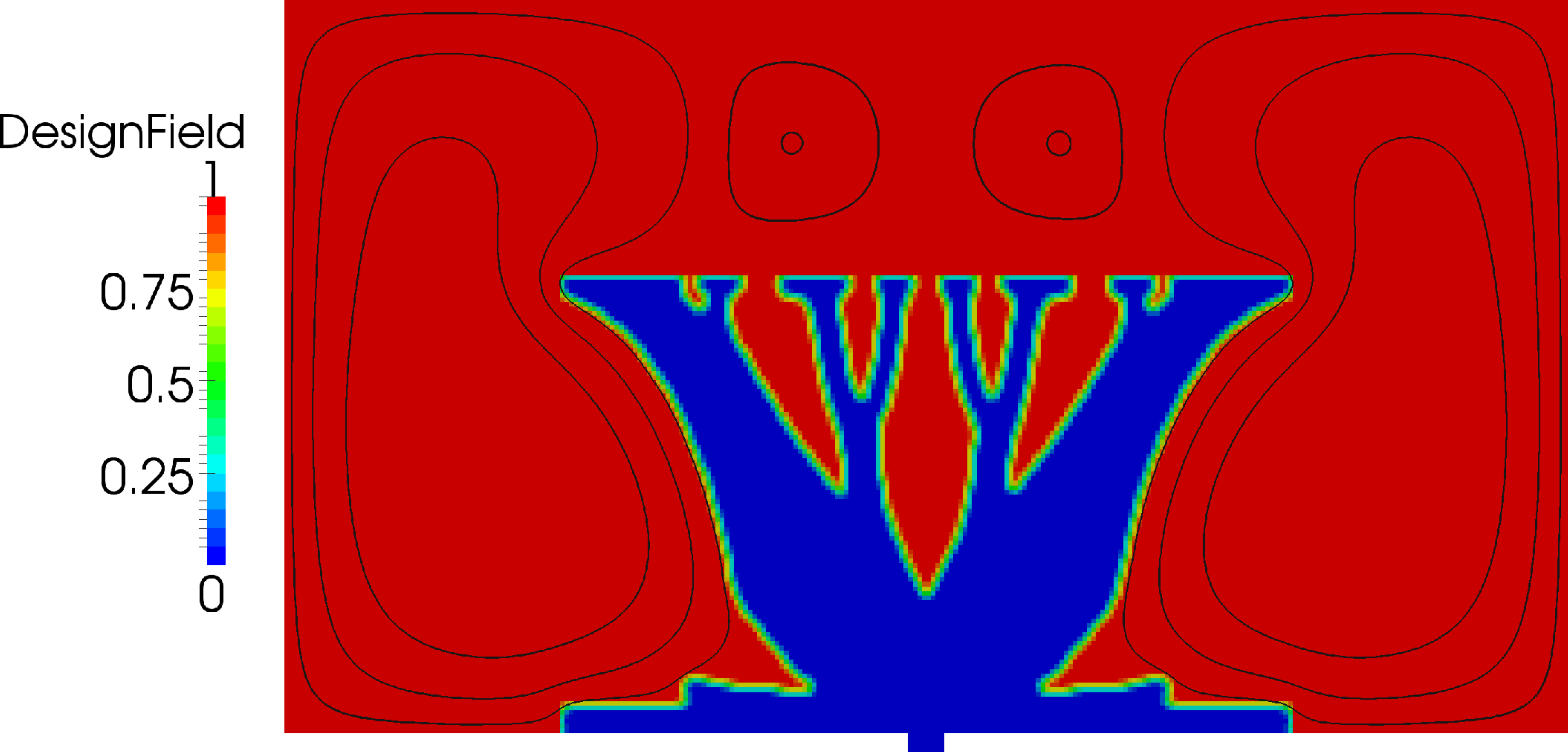}
\hspace*{0.025\textwidth}
\includegraphics[height=0.25\textwidth]{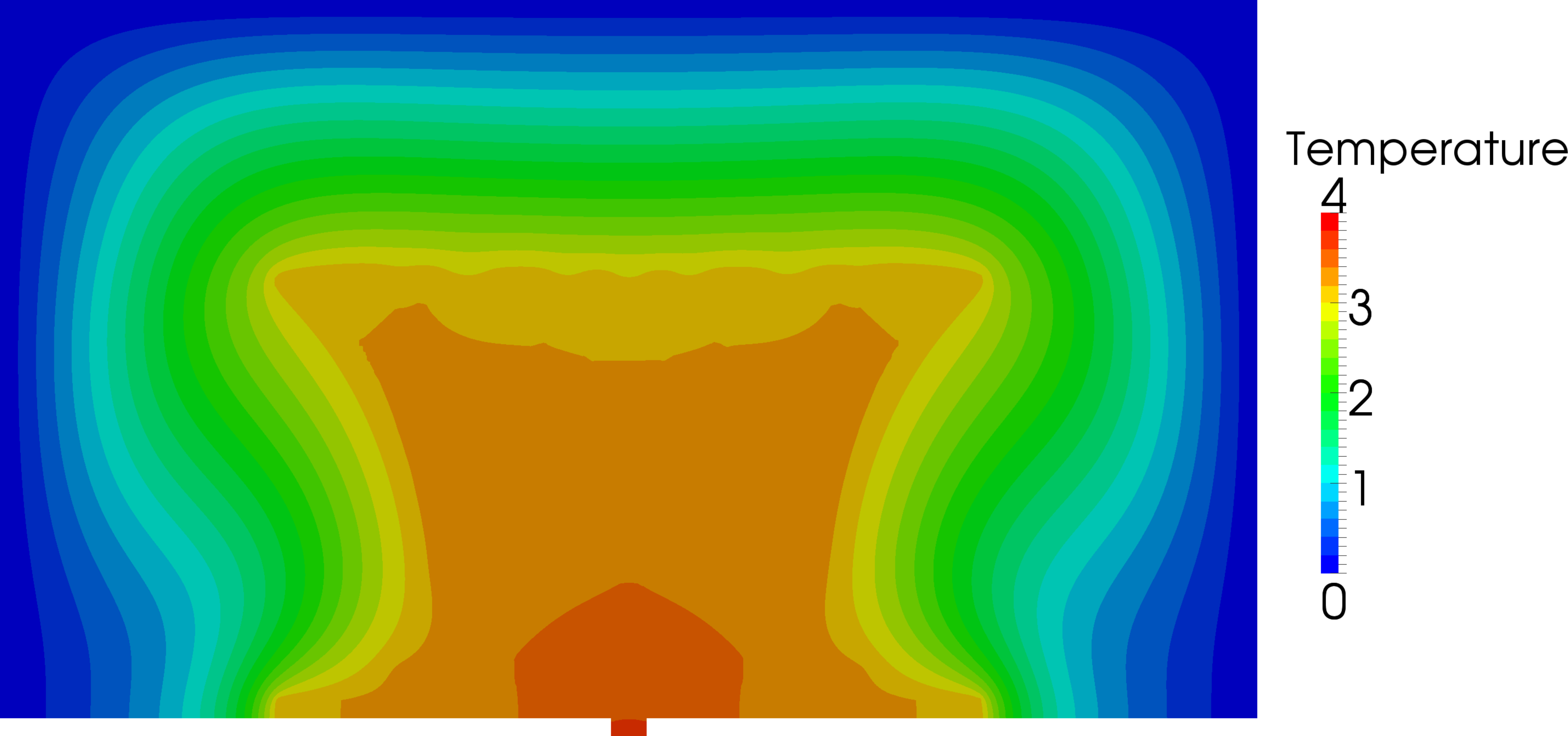}\label{fig:cpunatconv1_streamdes-b}}\\
\hspace*{-0.02\textwidth}\subfloat[$Gr_{H} = 6400$]{\hspace*{0.095\textwidth}\includegraphics[height=0.25\textwidth]{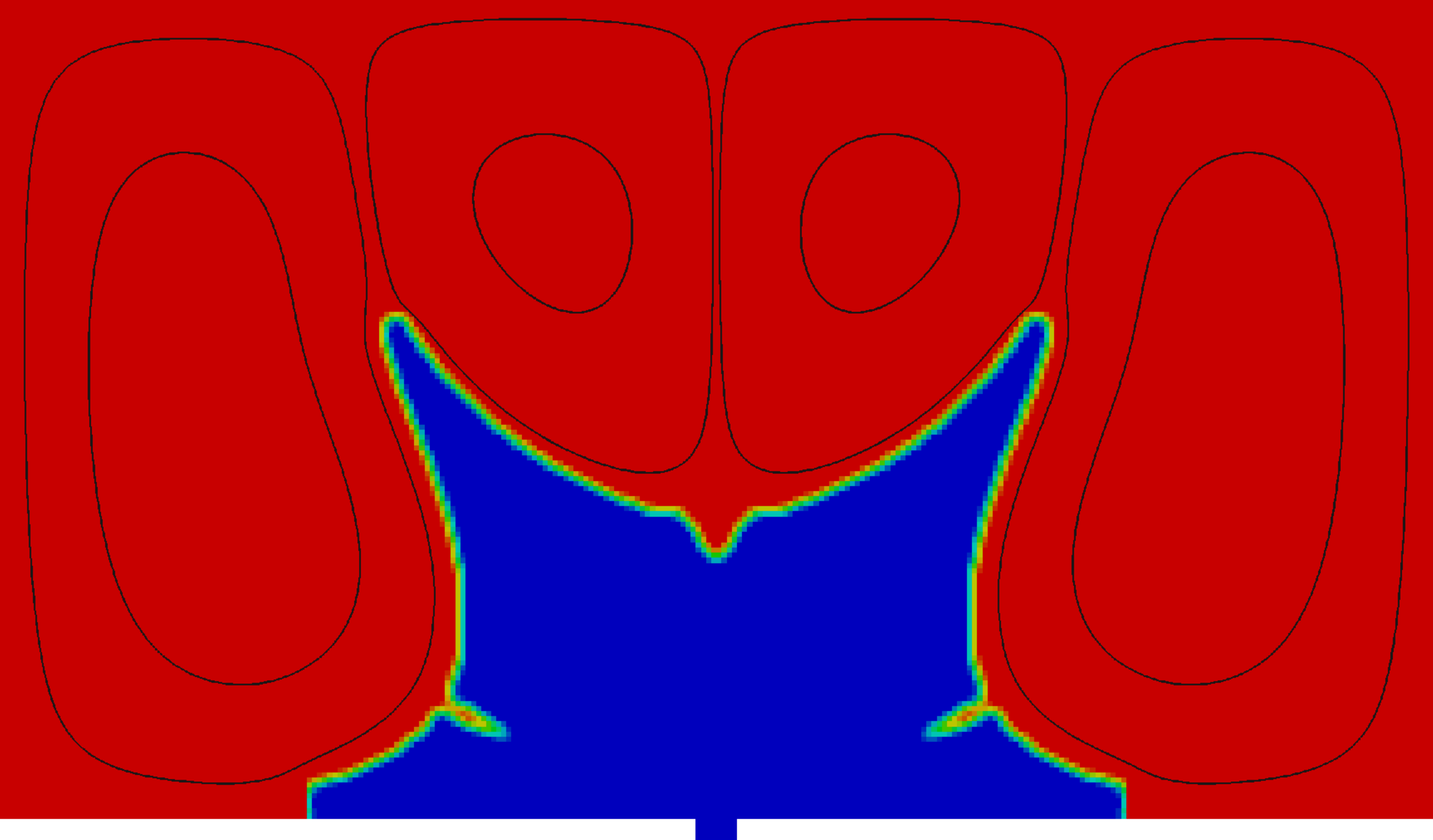}
\hspace*{0.025\textwidth}
\includegraphics[height=0.25\textwidth]{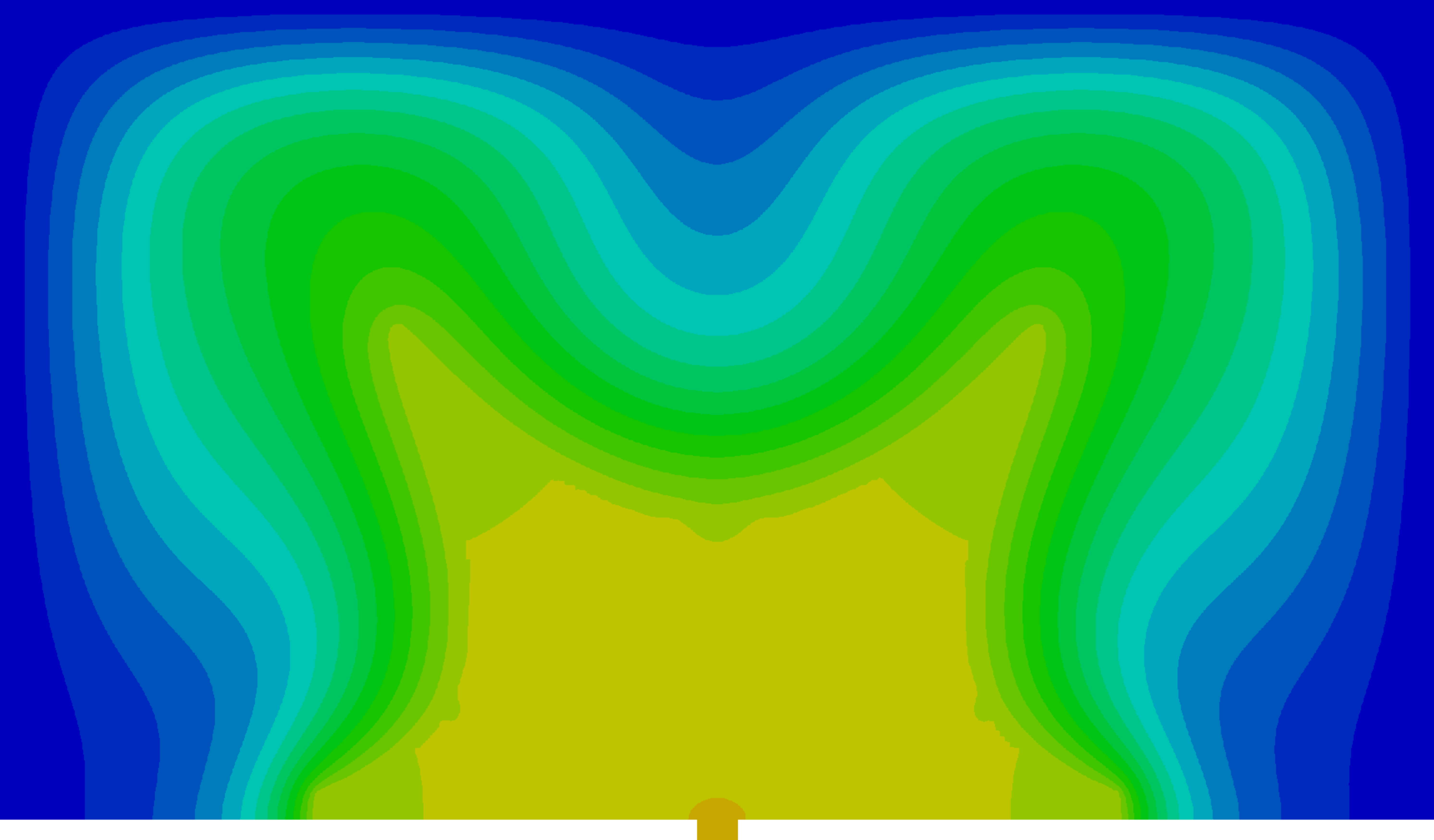}\hspace*{0.095\textwidth}\label{fig:cpunatconv1_streamdes-c}}
\caption{Optimised designs along with streamlines and corresponding temperature fields for the heat sink subjected to natural convection flow at various $Gr_{H}$. \\
Objective function: (a) $f_{tc} = 8.886\cdot10^{-4}$ - (b) $f_{tc} = 8.134\cdot10^{-4}$  - (c) $f_{tc} = 6.987\cdot10^{-4}$\\
Design iterations: (a) 440 - (b) 486 - (c) 376} \label{fig:cpunatconv1_streamdes}
\end{figure}
Figure \ref{fig:cpunatconv1_streamdes} shows the optimised designs for various $Gr_{H}$-numbers along with the streamlines illustrating the recirculatory convection rolls that form due to the natural convection effect. 
\begin{table}
\caption{Crosscheck objective function values ($\times 10^{-4}$) for the natural convection heat sink designs shown in figure \ref{fig:cpunatconv1_streamdes}.}\label{tab:cpunatconv1_crosscheck}
\centering
\tabsize
\begin{tabular}{cccc}
\toprule
 & \multicolumn{3}{c}{Optimisation $Gr_{H}$}\\
Analysis $Gr_{H}$ & $640$ & $3200$ & $6400$ \\
\midrule
$640$ & \textcolor{blue}{$8.886$} & $9.115$ & $10.52$\\
$3200$ & $8.587$ & \textcolor{blue}{$8.134$} & $8.461$ \\
$6400$ & $7.982$ & $7.271$ & \textcolor{blue}{$6.987$} \\
\bottomrule
\end{tabular}
\end{table}
Table \ref{tab:cpunatconv1_crosscheck} contains the crosscheck values for the different designs and flow conditions. For the crosscheck, the optimised designs are analysed across the different flow conditions and the design optimised for a certain flow condition should preferably perform better than the others for its particular flow condition. As can be seen from table \ref{tab:cpunatconv1_crosscheck}, the designs perform exactly as they should and one can therefore try to draw some conclusions from the obtained designs. It can clearly be seen from figure \ref{fig:cpunatconv1_streamdes} that significantly different designs are obtained for the different flow conditions.
For a Grashof number of 640, diffusion can be seen to dominate the heat transfer in the fluid when looking at the temperature field in the right of figure \ref{fig:cpunatconv1_streamdes-a}. This is clearly reflected in the obtained design, where the solid material is placed in the form of a conductive tree with branches conducting the heat towards the cold boundaries, similar to what is seen for pure conductive heat transfer problems \cite{Bendsoee2003,Wang2010}.
As the Grashof number is increased to 3200, as shown in figure \ref{fig:cpunatconv1_streamdes-b}, it can be seen that as convection begins to play a role in the heat transfer in the fluid, the obtained design begins to adapt to the fluid flow. It can be seen that no branches are formed towards the side walls in this case, but the two main branches have been thickened and shaped to accommodate the fluid motion.
These effects are exhibited even more clearly when the Grashof number is increased further to 6400. As can be seen from figure \ref{fig:cpunatconv1_streamdes-c}, the obtained design no longer shows significant branching and appears to have adapted the surface to optimise the contact with the four large convection cells by curving the edges.

One of the disadvantages of using the density filter is the transition region of intermediate relative densities that it inherently imposes along the interface between solid and fluid. It can be debated to what extent these represent a problem for the accuracy of the modelling of the actual physics and whether they impose difficulties for interpreting the obtained designs. The lack of a distinct and sharply defined interface between the solid and fluid regions could be fixed by using projection filters or robust optimisation techniques \cite{Sigmund2007,Wang2010}, but these introduce additional optimisation parameters which would require tuning for the problems at hand. As long as simple thresholding of the relative densities produces physically meaningful and feasible designs, then it is not seen as pertinent to implement projection filters.

\begin{figure}
\centering
\hspace*{-0.045\textwidth}\subfloat[$Gr_{H} = 640$]{\hspace*{0.12\textwidth}\includegraphics[height=0.25\textwidth]{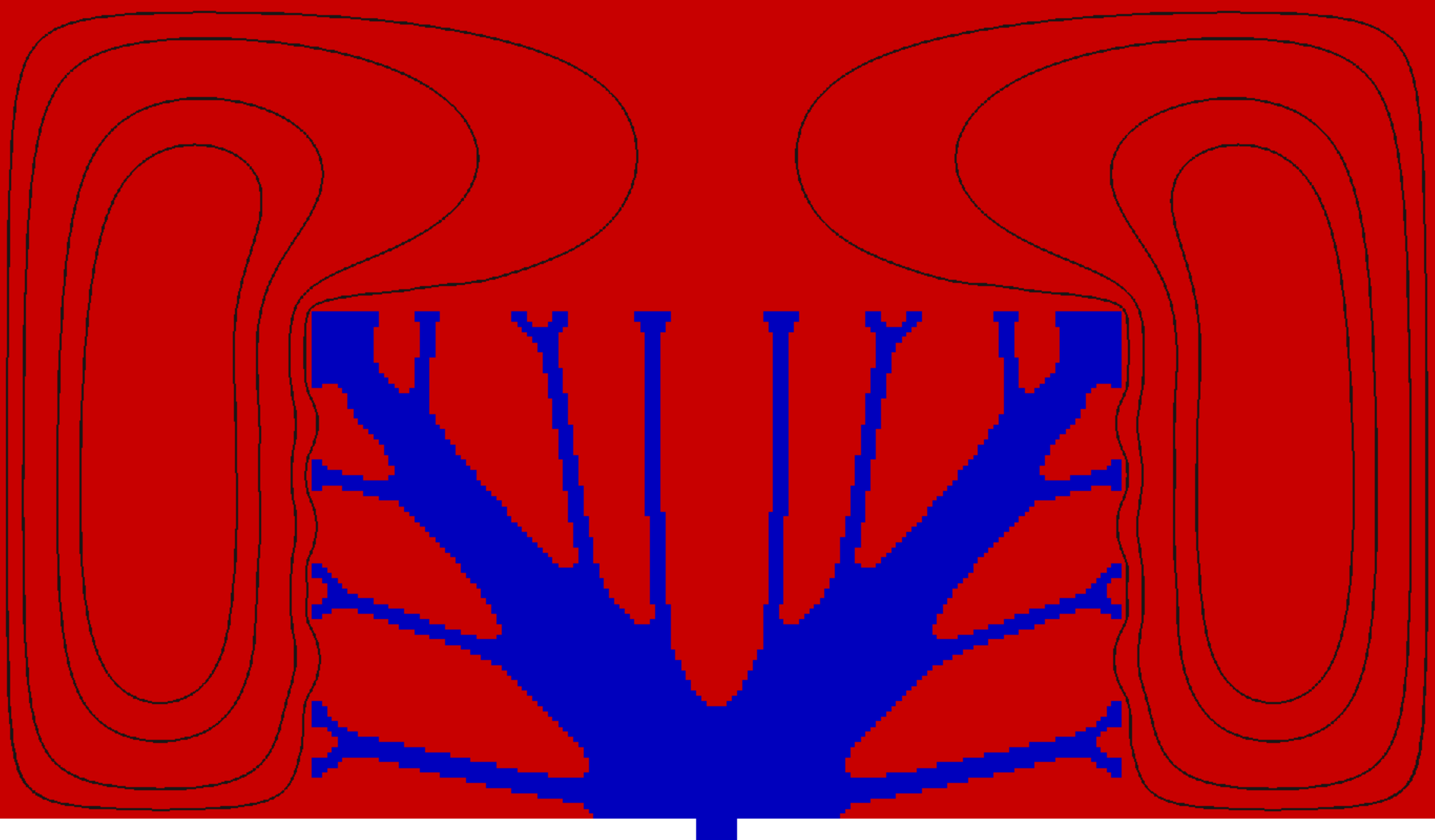}
\hspace*{0.025\textwidth}
\includegraphics[height=0.25\textwidth]{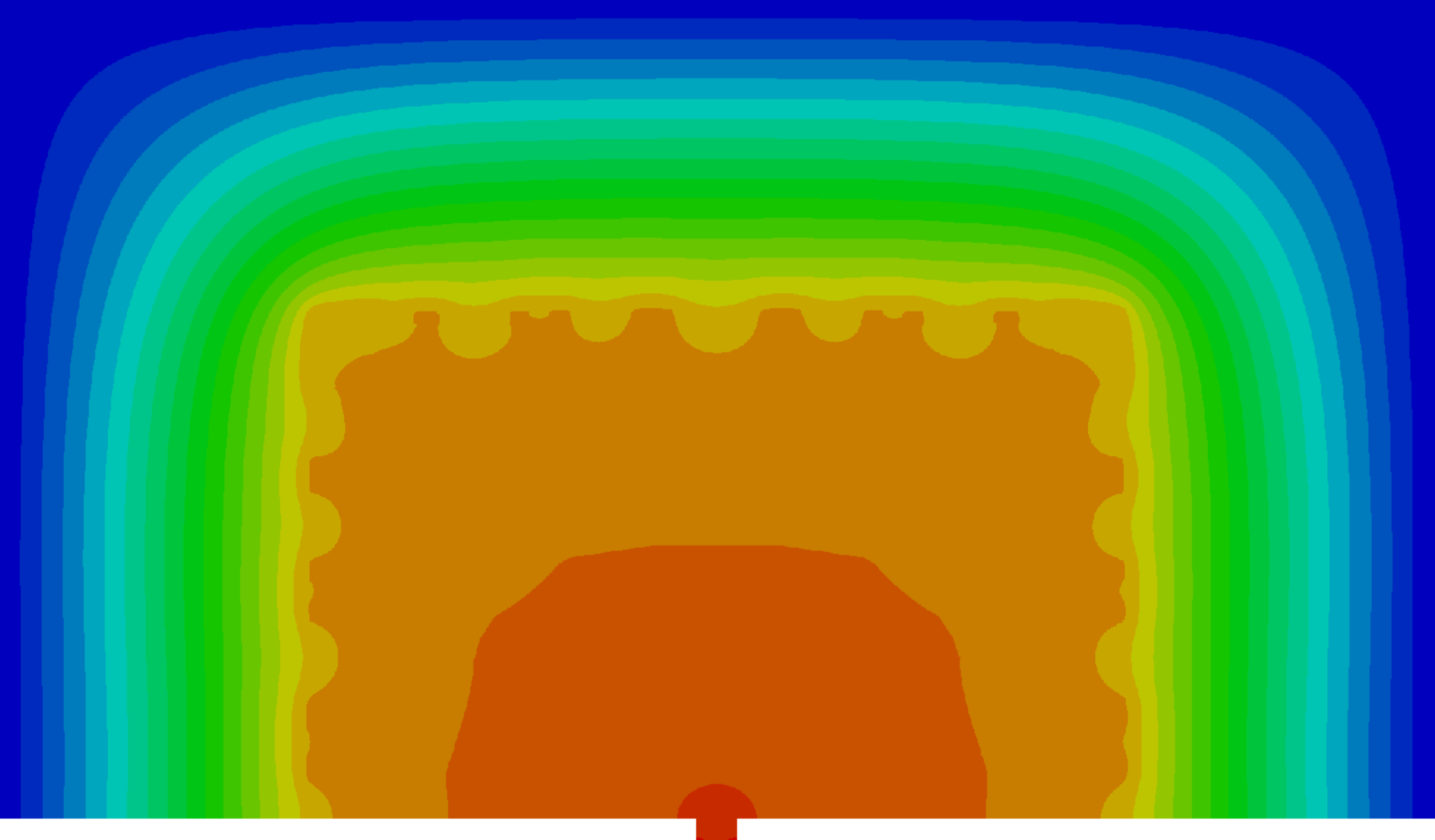}
\hspace*{0.12\textwidth}\label{fig:cpunatconv1_thresh01_streamdes-a}}\\
\hspace*{-0.045\textwidth}\subfloat[$Gr_{H} = 3200$]{\includegraphics[height=0.25\textwidth]{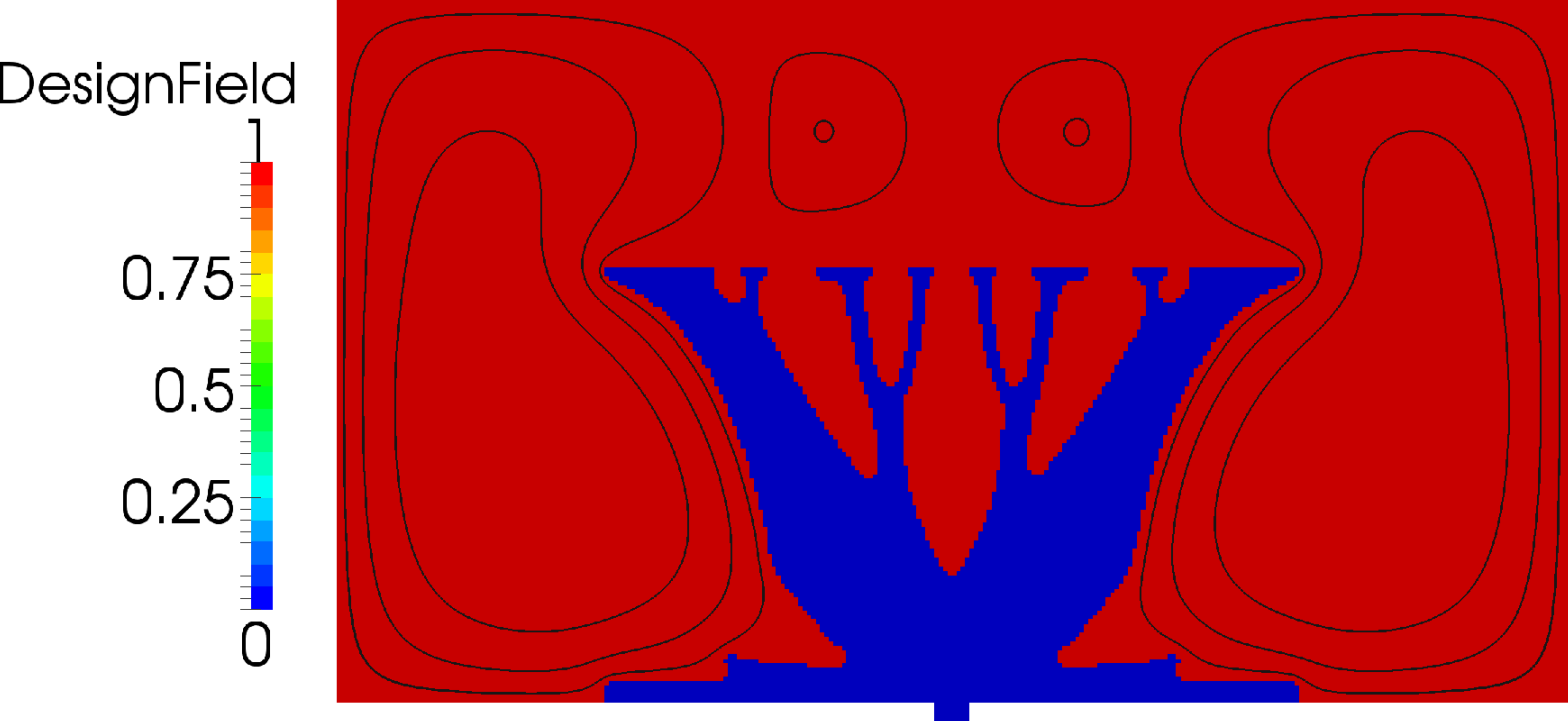}
\hspace*{0.025\textwidth}
\includegraphics[height=0.25\textwidth]{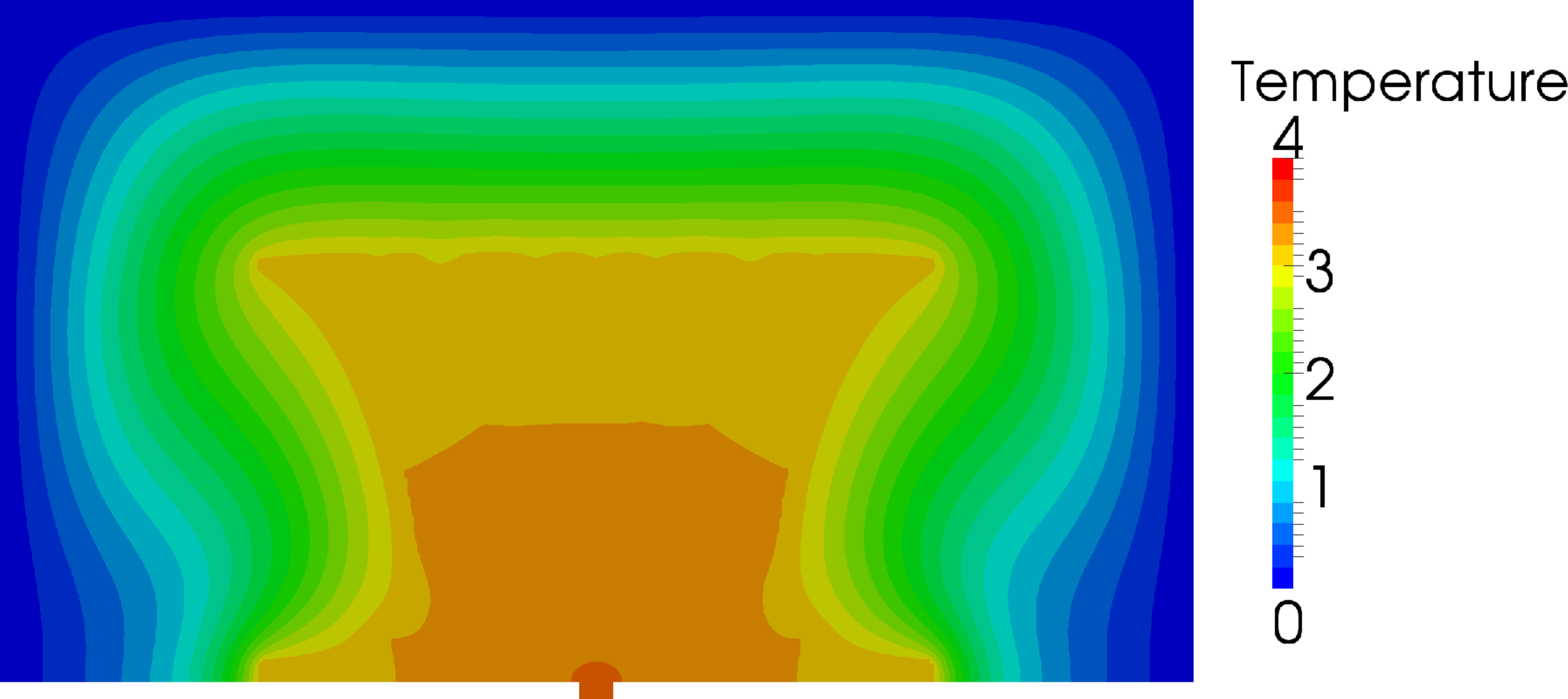}
\label{fig:cpunatconv1_thresh01_streamdes-b}}\\
\hspace*{-0.045\textwidth}\subfloat[$Gr_{H} = 6400$]{\hspace*{0.12\textwidth}\includegraphics[height=0.25\textwidth]{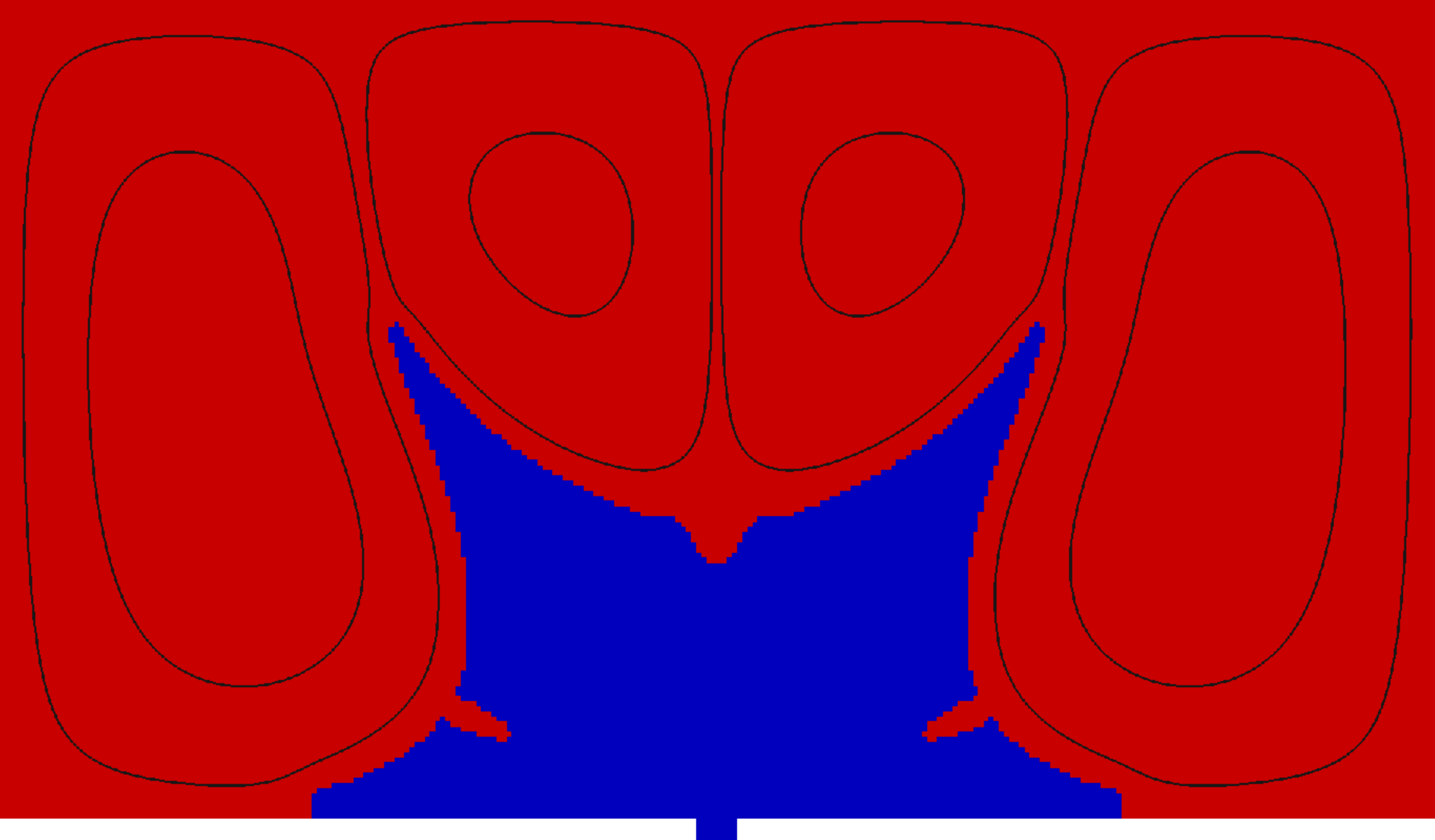}
\hspace*{0.025\textwidth}
\includegraphics[height=0.25\textwidth]{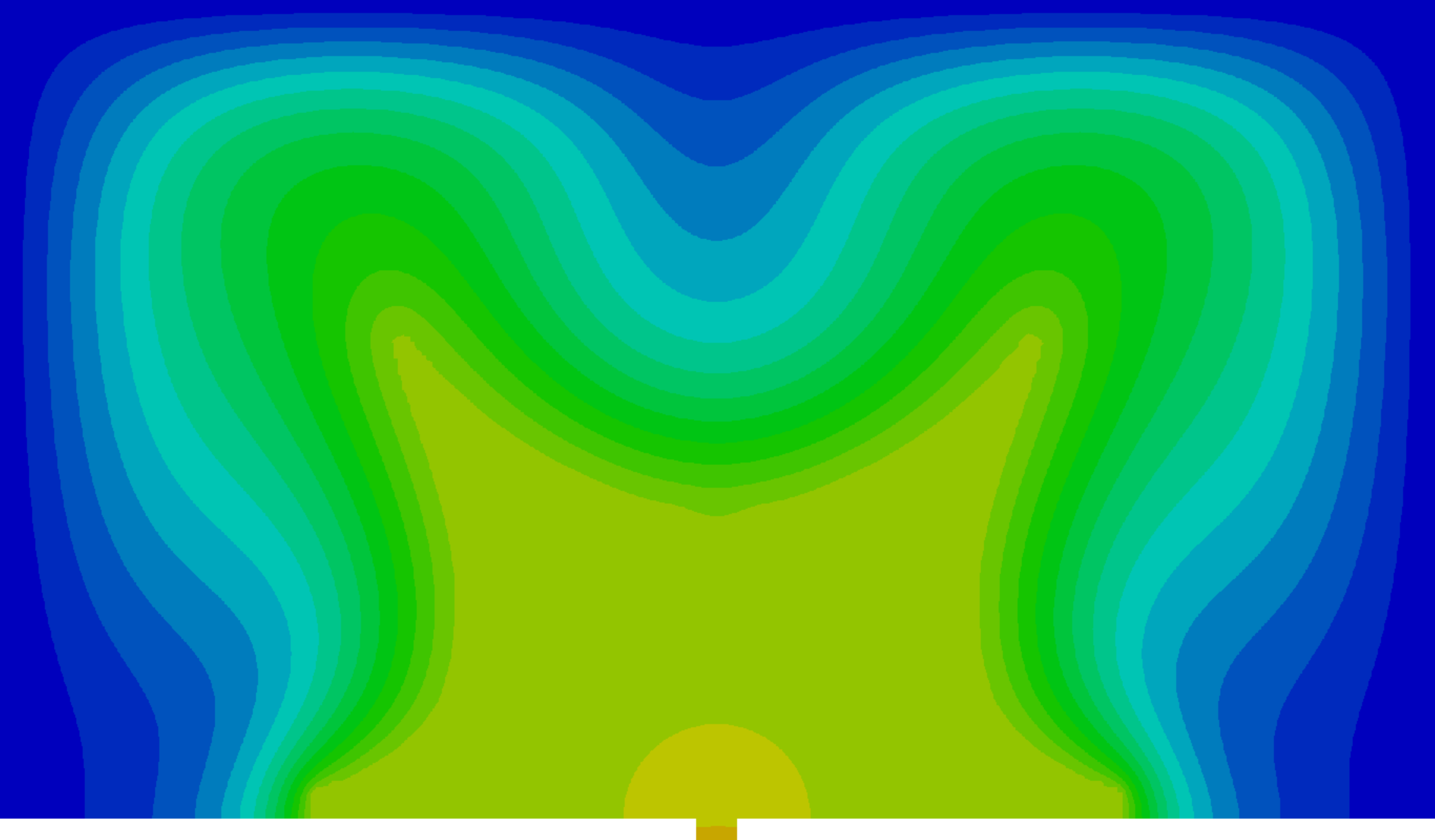}
\hspace*{0.12\textwidth}\label{fig:cpunatconv1_thresh01_streamdes-c}}\\
\caption{Thresholded versions, of the optimised designs in figure \ref{fig:cpunatconv1_streamdes}, along with streamlines and corresponding temperature fields for the heat sink subjected to natural convection flow at various $Gr_{H}$. \\
Objective function: (a) $f_{tc} = 8.369\cdot10^{-4}$ - (b) $f_{tc} = 7.882\cdot10^{-4}$  - (c) $f_{tc} = 6.753\cdot10^{-4}$} \label{fig:cpunatconv1_thresh01_streamdes}
\end{figure}
As the physical parameters, the impermeability and the effective conductivity, are penalised quite aggressively at the final stages of the optimisation, it seems reasonable to set the threshold value at $\gamma_{th} = 0.1$; setting all relative densities below to 0 and all above to 1. This is deemed to be reasonable as the thresholding procedure leads to the binary designs shown in figure \ref{fig:cpunatconv1_thresh01_streamdes} that perform as seen in table \ref{tab:cpunatconv1_thresh01_crosscheck}.
\begin{table}
\caption{Crosscheck objective function values ($\times 10^{-4}$) for the thresholded natural convection heat sink designs shown in figure \ref{fig:cpunatconv1_thresh01_streamdes}.}
\label{tab:cpunatconv1_thresh01_crosscheck}
\centering
\tabsize
\begin{tabular}{cccc}
\toprule
 & \multicolumn{3}{c}{Optimisation $Gr_{H}$}\\
Analysis $Gr_{H}$ & $640$ & $3200$ & $6400$ \\
\midrule
$640$ & \textcolor{blue}{$8.369$} & $8.828$ & $10.29$\\
$3200$ & $8.082$ & \textcolor{blue}{$7.882$} & $8.200$ \\
$6400$ & $7.495$ & $7.040$ & \textcolor{blue}{$6.753$} \\
\bottomrule
\end{tabular}
\end{table}
The crosscheck shows that the threshold procedure leads to designs that still perform as they should with respect to the other flow conditions. Comparing tables \ref{tab:cpunatconv1_crosscheck} and \ref{tab:cpunatconv1_thresh01_crosscheck}, it can also be seen that the threshold procedure actually improves the objective functional values as compared to the original design distributions. This can be attributed to the fact that penalised solid material, with a relative density between 0 and 0.1 but a low effective conductivity, is ``upgraded'' to fully conducting solid material due to the threshold. It should be noted that the maximum allowable solid volume fraction, which was imposed as an optimisation constraint, is not exceeded for all three thresholded designs, with used solid volume fractions of $38.6\%$, $42.9\%$ and $46.4\%$ for $Gr_{H} = \left\lbrace 640, 3200, 6400 \right\rbrace$ respectively. Thus, the thresholded designs shown in figure \ref{fig:cpunatconv1_thresh01_streamdes} are seen as physically meaningful, feasible and representable designs for the optimisation problem.

\subsection{Natural convection pump}

\begin{figure}
\centering
\includegraphics[height=0.4\textwidth]{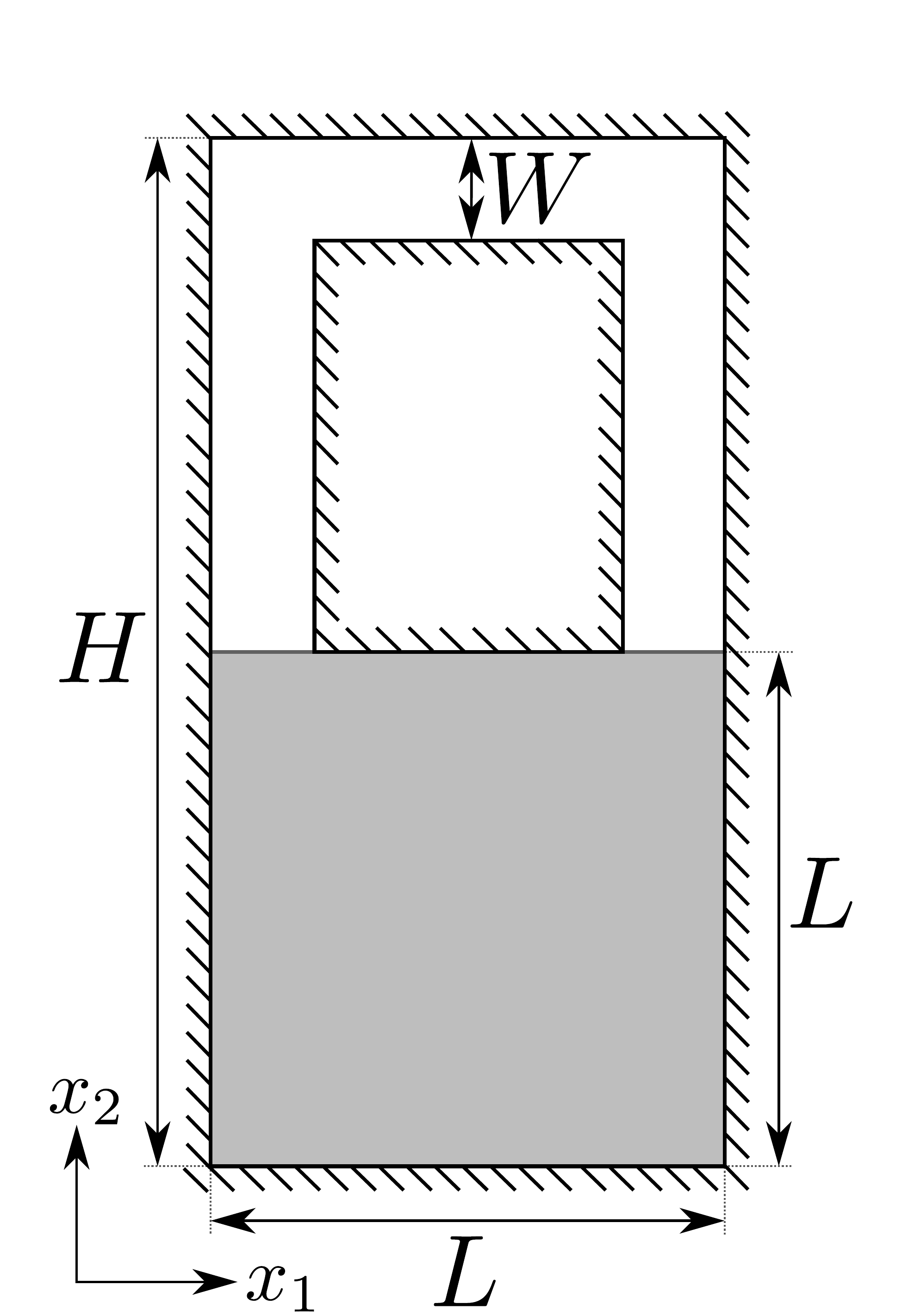}\hspace*{0.05\textwidth}
\includegraphics[height=0.4\textwidth]{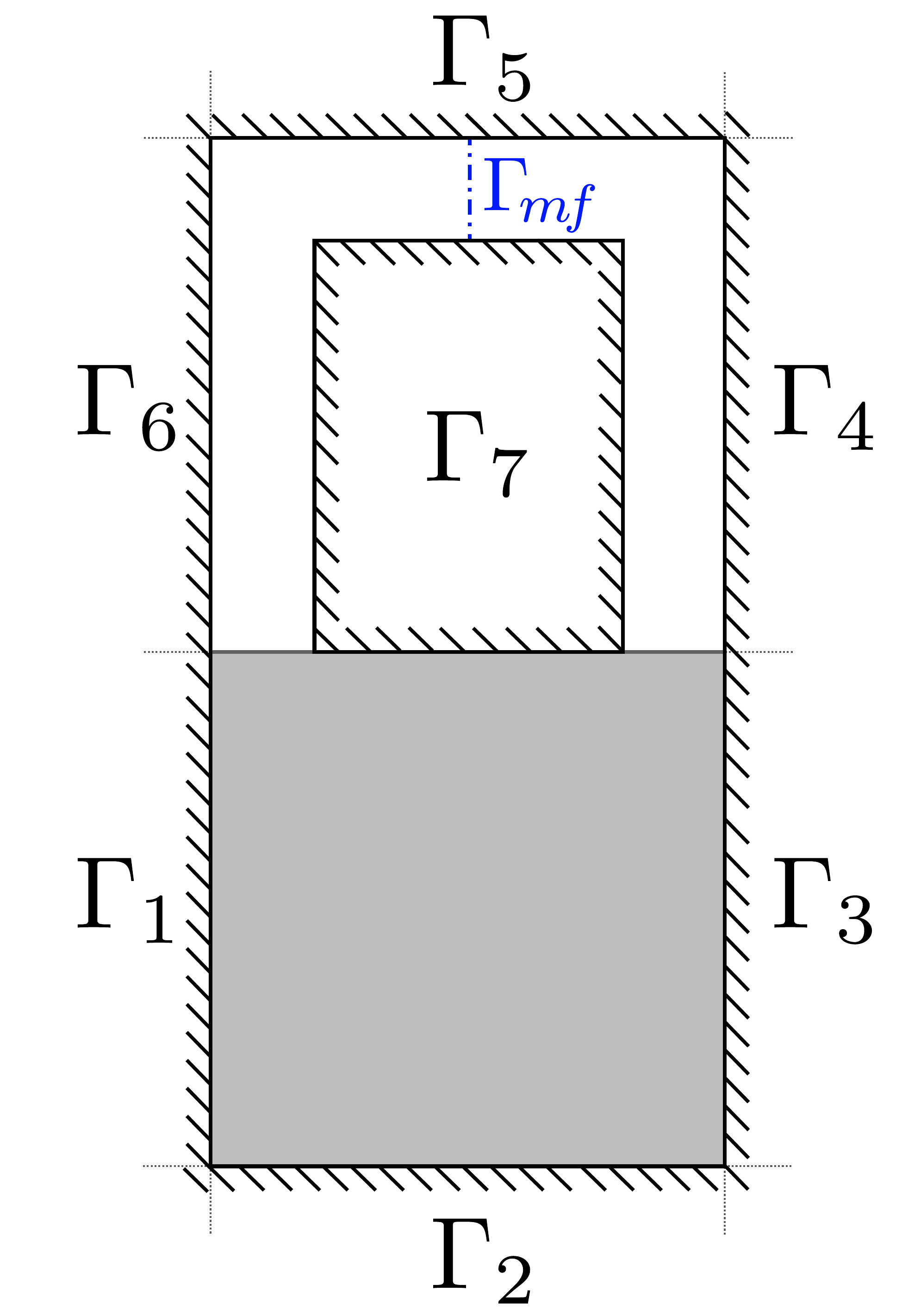}
\includegraphics[height=0.4\textwidth]{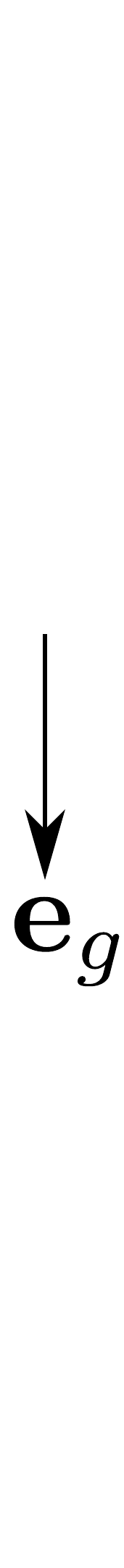}
\caption{Schematic illustration of the layout and boundary conditions for the natural convection micropump problem. White denotes fully fluid and light grey denotes the design domain.} \label{fig:natconvpump_schematic}
\end{figure}
The second numerical example is the design of a micropump where the fluid motion is caused by natural convection due to differential heating of walls. Figure \ref{fig:natconvpump_schematic} shows schematic illustrations of the layout and boundary conditions for the problem. The calculation domain consists of a square design domain, which is connected to itself through a closed-loop channel system.
\begin{table}
\caption{The dimensionless quantities used for the natural convection micropump problem shown in figure \ref{fig:natconvpump_schematic}.}\label{tab:natconvpump_sizes}
\centering
\tabsize
\begin{tabular}{ccccc}
\toprule
Total height & Design size & Channel width \\
\midrule
$H=2$ & $L=1$ & $W=0.2$\\
\bottomrule
\end{tabular}
\end{table}
Table \ref{tab:natconvpump_sizes} lists the dimensionless quantities specifying the layout of the natural convection micropump problem.
All of the quantities specified are kept constant throughout. All spatial dimensions are relative to the height and width of the design domain, $L=1$, the flow velocities are relative to the diffusion velocity as explained in section \ref{sec:govequ_dimless} and the temperatures are relative to a reference temperature difference, $\Delta T$.

The problem is investigated for two different orientations, with respect to the gravitational direction, and several combinations of boundary conditions.
\begin{table}
\caption{The different combinations of orientation and boundary conditions used for the natural convection micropump problem shown in figure \ref{fig:natconvpump_schematic}.}\label{tab:natconvpump_bcs}
\centering
\tabsize
\begin{tabular}{ccccccccc}
\toprule
Number & Orientation & $\Gamma_{1}$ & $\Gamma_{2}$ & $\Gamma_{3}$ & $\Gamma_{4}$ & $\Gamma_{5}$ & $\Gamma_{6}$ & $\Gamma_{7}$ \\
\midrule
1 & Vertical & $T_{1} = 1$ & $f_{n} = 0$ & $T_{3} = 0$ & $f_{n} = 0$ & $f_{n} = 0$ & $f_{n} = 0$ & $f_{n} = 0$\\
2 & Vertical & $f_{n} = 0$ & $T_{2} = 1$ & $f_{n} = 0$ & $f_{n} = 0$ & $T_{5} = 0$ & $f_{n} = 0$ & $f_{n} = 0$\\
3 & Horizontal & $T_{1} = 1$ & $f_{n} = 0$ & $T_{3} = 0$ & $f_{n} = 0$ & $f_{n} = 0$ & $f_{n} = 0$ & $f_{n} = 0$\\
\bottomrule
\end{tabular}
\end{table}
Table \ref{tab:natconvpump_bcs} lists the combinations of boundary conditions and configurations used for the natural convection micropump problem.
The problem is studied under constant parameters, $Gr_{L}=10^{3}$, $C_{k}=10^{-2}$, $\overline{\alpha}=10^{6}$ and $Pr=1$. 

The design domain is discretised using $50\times50$ square elements and the closed-loop channel system is discretised using elements of the same size, that is $10$ elements over the width, making a total of 3800 elements. The total number of state degrees of freedom is $16,960$ for the entire calculation domain.  The density filter is not applied for this problem.

The objective functional for the micropump problem is chosen as the mass flow through the surface $\Gamma_{mf}$ indicated by the blue line in figure \ref{fig:natconvpump_schematic} and the direction of the mass flow can be chosen by either maximising or minimising the mass flow functional:
\begin{equation}
f_{m\! f} ( \vecrsym{\gamma}, \vecr{t} ) = \int_{\Gamma_{\! m\! f}} \vecr{u}^{T} \vecr{n} \,dS
\end{equation}
where $\vecr{u}$ is the vector containing the nodal velocities and $\vecr{n}$ is the surface normal vector.
For the micropump problem, a constraint on the fluid volume fraction is imposed instead of on the solid volume fraction. Even though it is not necessarily beneficial to have a completely fluid domain, the best results have been obtained by imposing a maximum allowable fluid volume fraction of $50\%$.

\begin{figure}
\centering
\includegraphics[height=0.3\textwidth]{gravity.pdf}
\subfloat[Design field]{\includegraphics[height=0.3\textwidth]{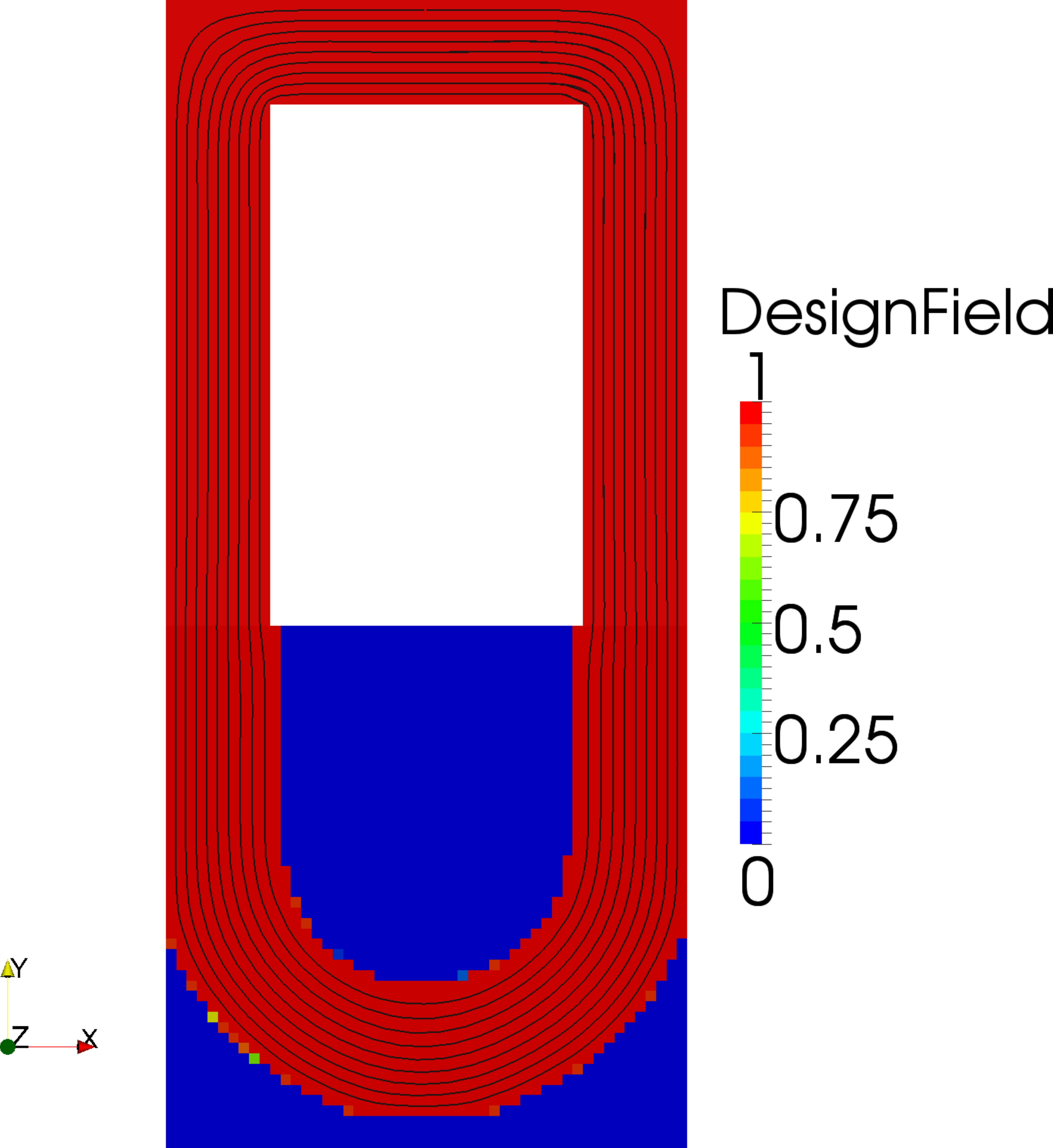}
\label{fig:cpunatconv1_contin3-a}}
\subfloat[Temperature field]{\includegraphics[height=0.3\textwidth]{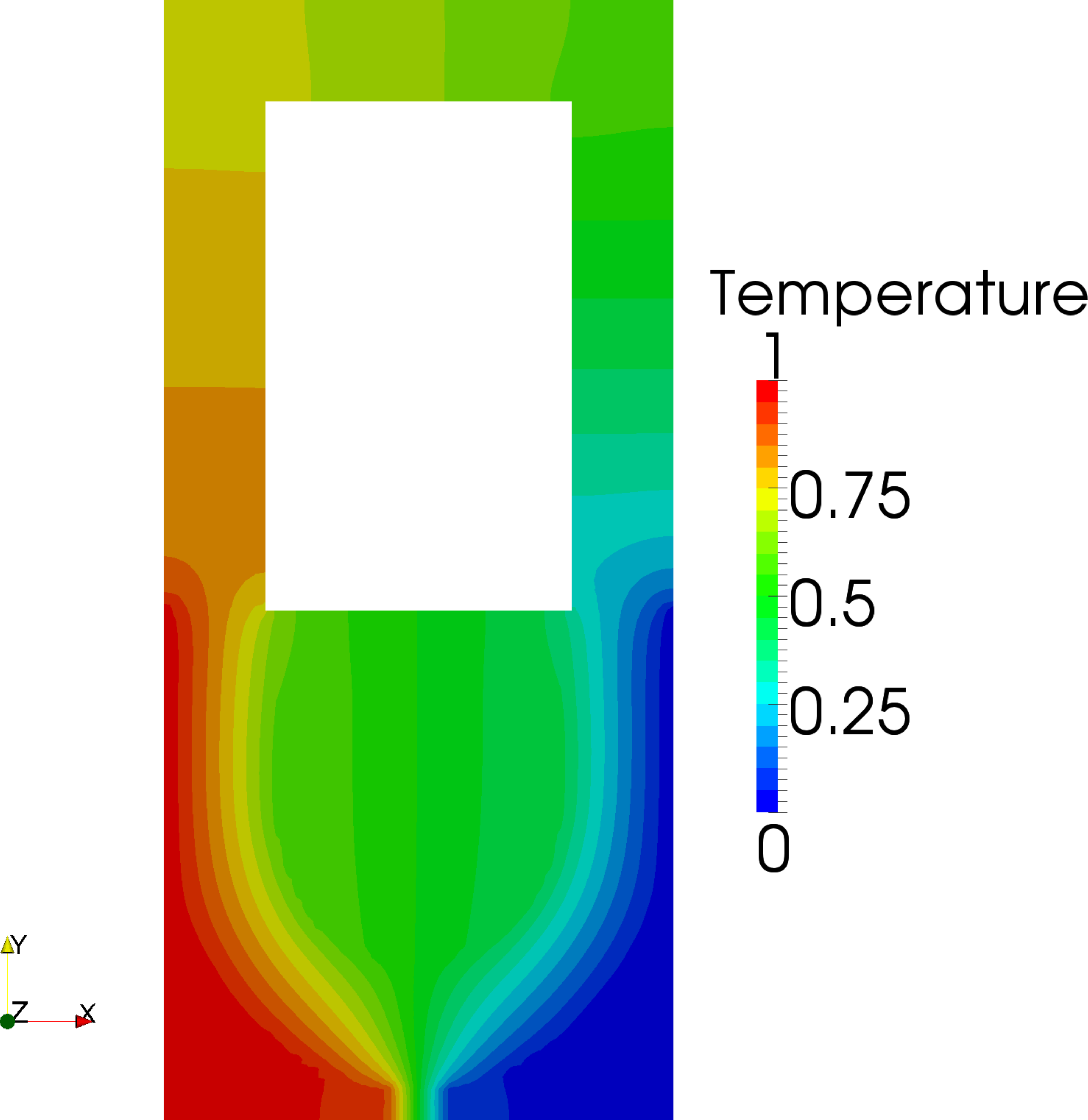}
\label{fig:cpunatconv1_contin3-b}}
\subfloat[Velocity field]{\includegraphics[height=0.3\textwidth]{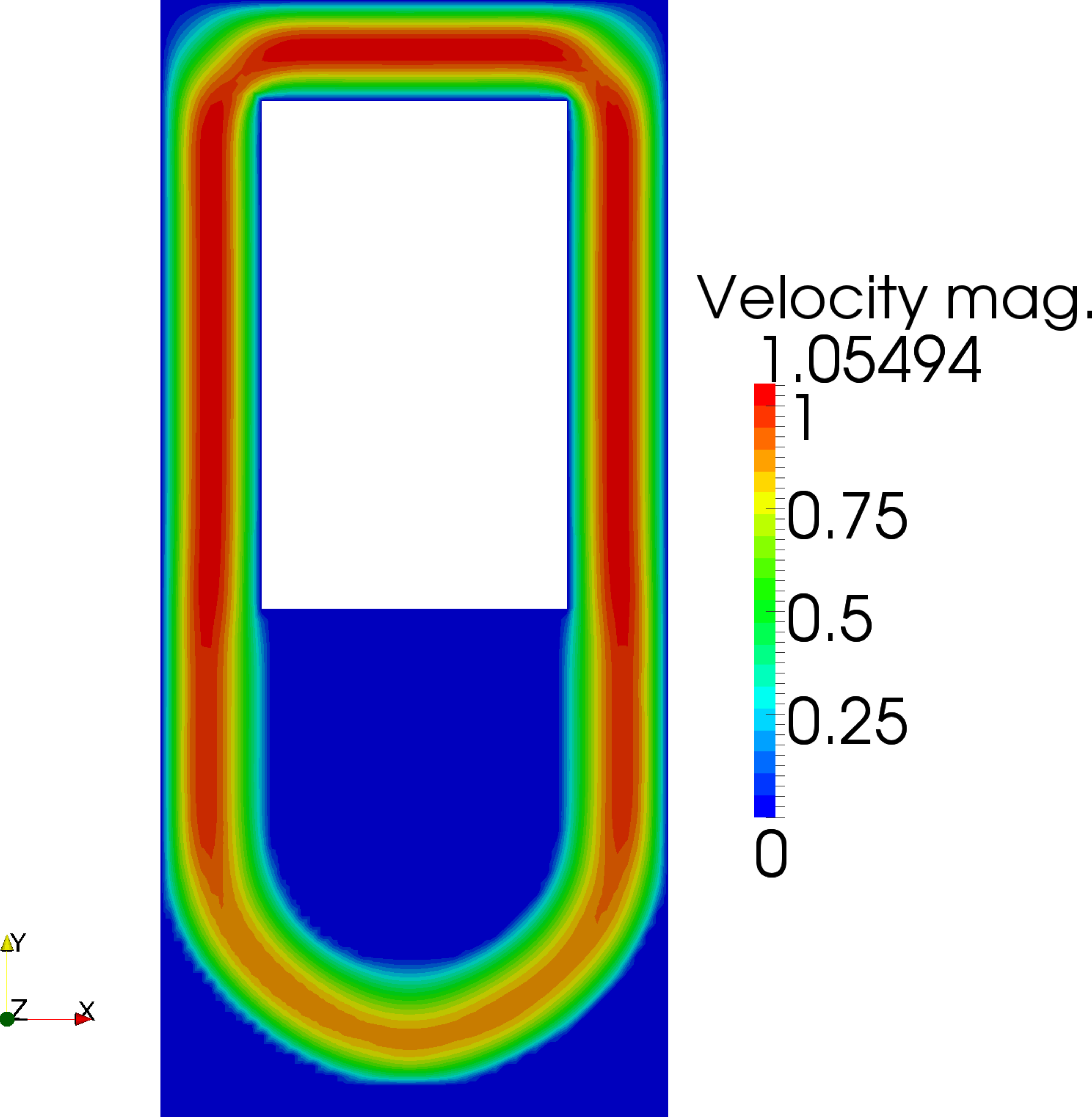}
\label{fig:cpunatconv1_contin3-c}}
\caption{Optimised design along with streamlines and corresponding temperature and velocity fields for clockwise massflow for combination 1 of the natural convection micropump. \\
Objective function: $f_{m\! f} = -1.390\cdot10^{-4}$ - Design iterations: 261} \label{fig:cpunatconv1_contin3}
\end{figure}
Figure \ref{fig:cpunatconv1_contin3} shows the optimised design along with the temperature and velocity fields for the first boundary condition combination listed in table \ref{tab:natconvpump_bcs}.  The obtained design is qualitatively symmetric about the vertical midplane and the temperature field is anti-symmetric about the same plane. A curious detail of the design can be seen from looking at the bottom of the temperature field in figure \ref{fig:cpunatconv1_contin3-b}. It can be seen that there is a very high temperature gradient over a relatively short span and this is due to the relative densities not quite being 0 in this region, yielding a significantly lower effective conductivity due to the convexity parameter being set to $q_{f} = 10^{4}$, however, this was needed in order to obtain a physical design solution for this problem. This is a problem with proper interpolation and will be discussed in section \ref{sec:discconc}.

\begin{figure}
\centering
\includegraphics[height=0.3\textwidth]{gravity.pdf}
\subfloat[Design field]{\includegraphics[height=0.3\textwidth]{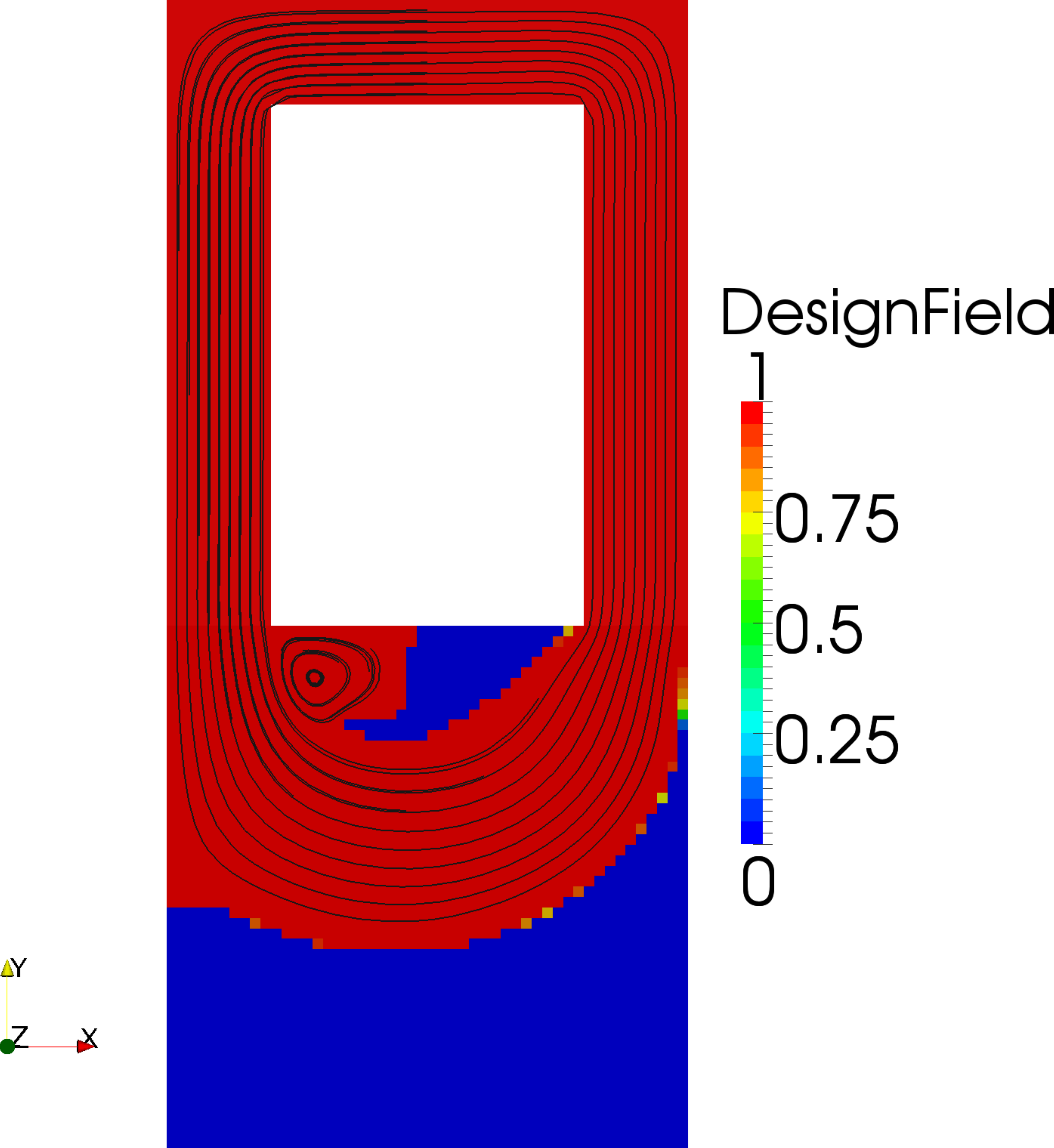}
\label{fig:cpunatconv1TBF_contin5-a}}
\subfloat[Temperature field]{\includegraphics[height=0.3\textwidth]{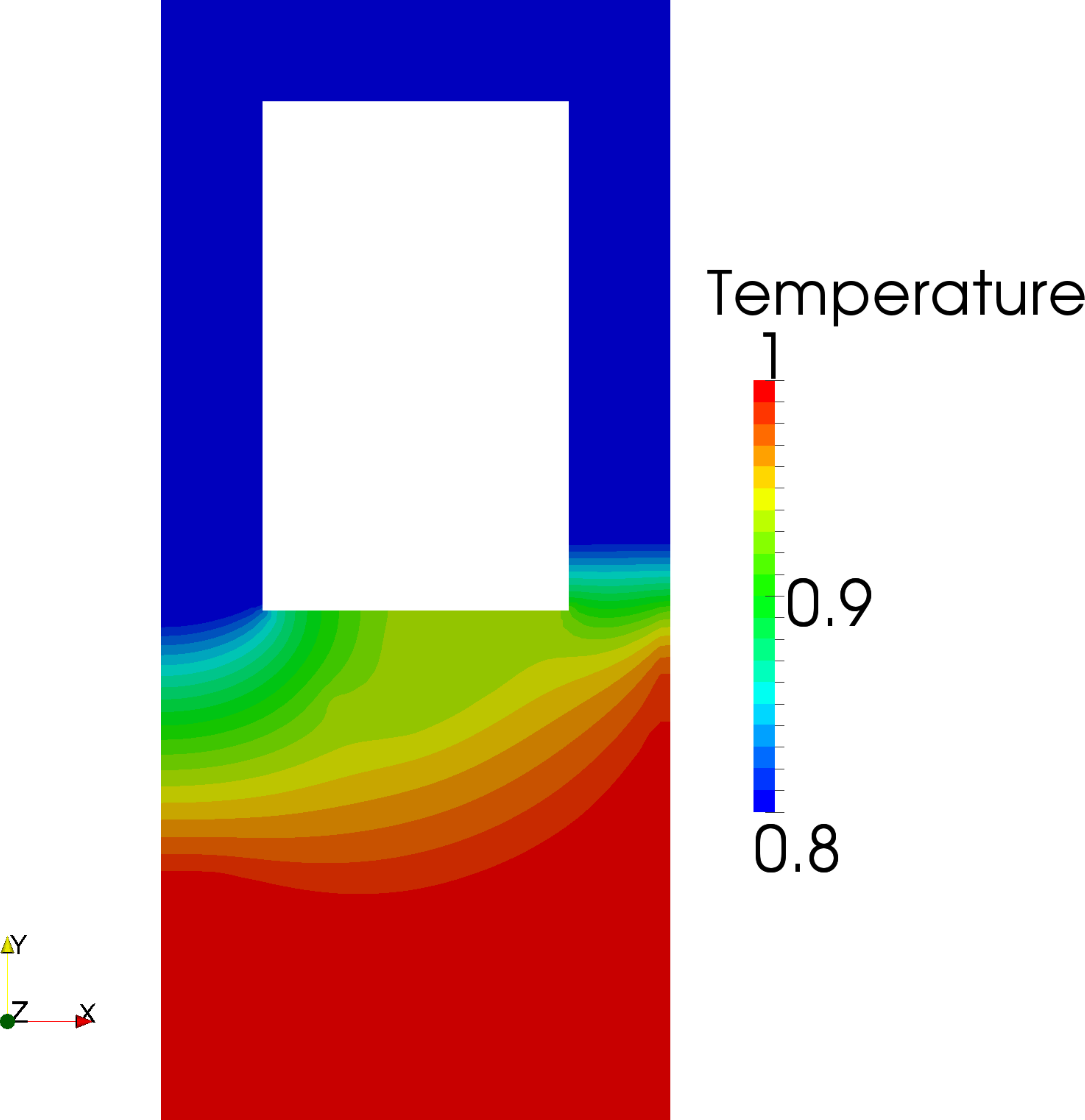}
\label{fig:cpunatconv1TBF_contin5-b}}
\subfloat[Velocity field]{\includegraphics[height=0.3\textwidth]{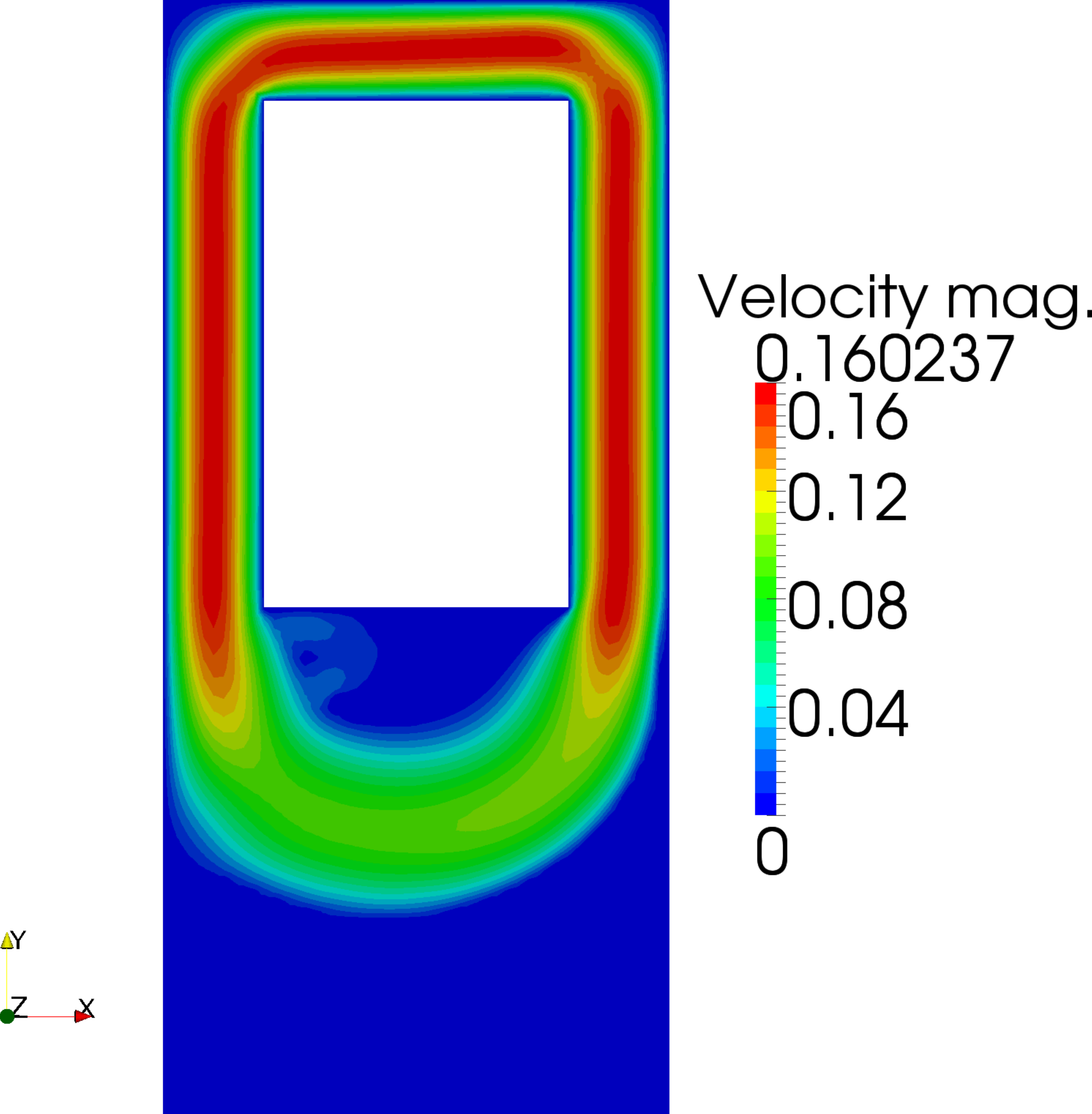}
\label{fig:cpunatconv1TBF_contin5-c}}
\caption{Optimised design along with streamlines and corresponding temperature and velocity fields for counter-clockwise massflow for combination 2 of the natural convection micropump. \\
Objective function: $f_{m\! f} = -2.090\cdot10^{-5}$ - Design iterations: 260} \label{fig:cpunatconv1TBF_contin5}
\end{figure}
Figure \ref{fig:cpunatconv1TBF_contin5} shows the optimised design along with the temperature and velocity fields for the second boundary condition combination listed in table \ref{tab:natconvpump_bcs}. 
It can be seen that a non-symmetric design is obtained and this physically makes sense. If one were to analyse the problem for this combination of boundary conditions with only fluid in the design domain, there would be no circulation through the closed-loop channel system due to the temperature distribution, and thus the buoyancy forces, being symmetric about the vertical midplane. In order to get circulation through the channel system, a non-symmetric feature or disturbance needs to be introduced. This means that one can also optimise the same problem for maximising the mass flow in the opposite direction. As expected, this yields almost the same design but mirrored across the vertical midplane and is thus not shown here.

\begin{figure}
\centering
\hspace*{-0.5cm}
\includegraphics[height=0.2\textwidth]{gravity.pdf}
\subfloat[Design field]{\includegraphics[height=0.2\textwidth]{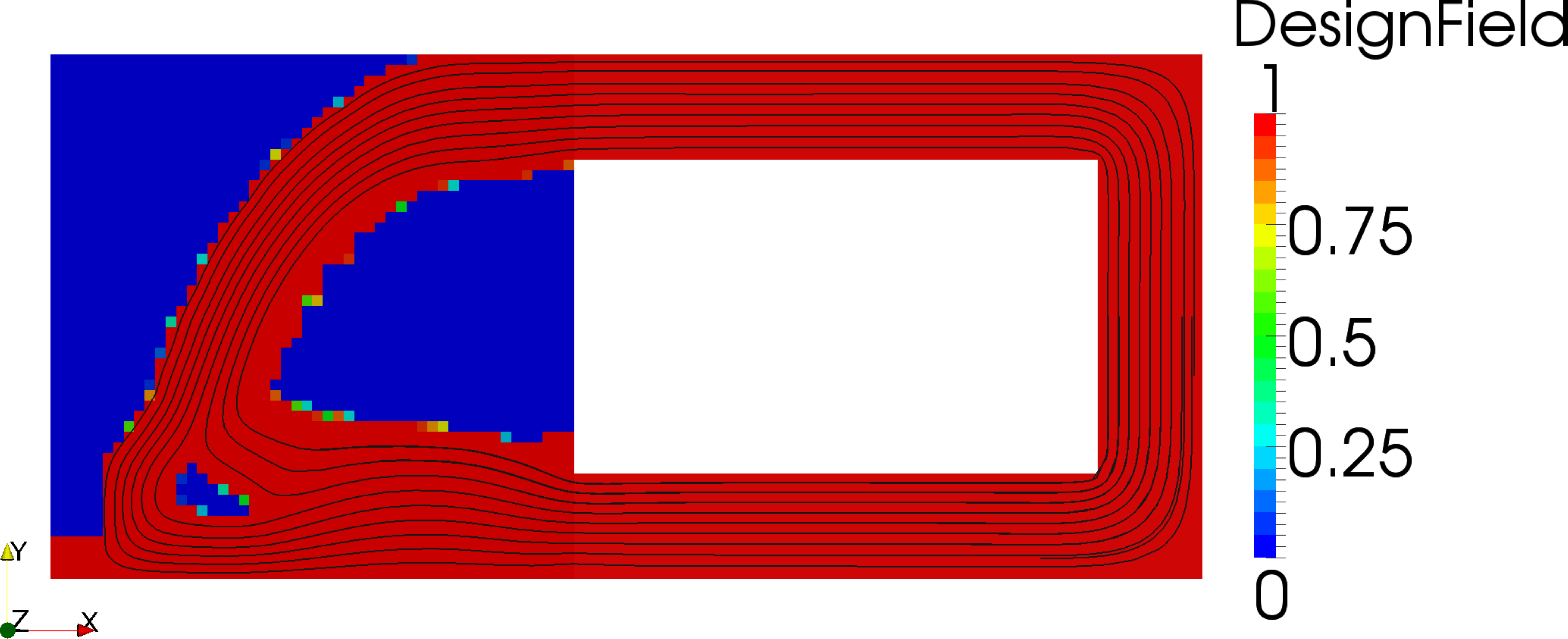}
\label{fig:cpunatconv2_contin0-a}}\\
\subfloat[Temperature field]{\includegraphics[height=0.2\textwidth]{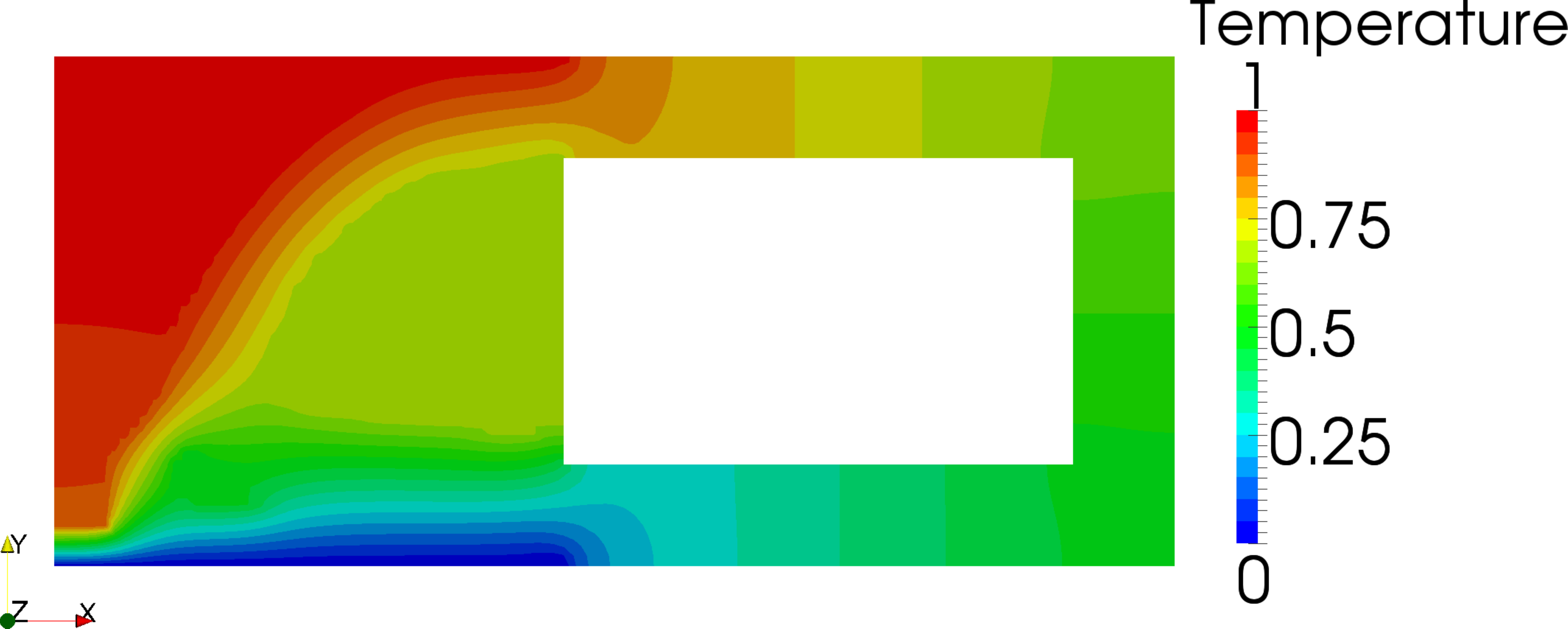}
\label{fig:cpunatconv2_contin0-b}}\\
\subfloat[Velocity field]{\includegraphics[height=0.2\textwidth]{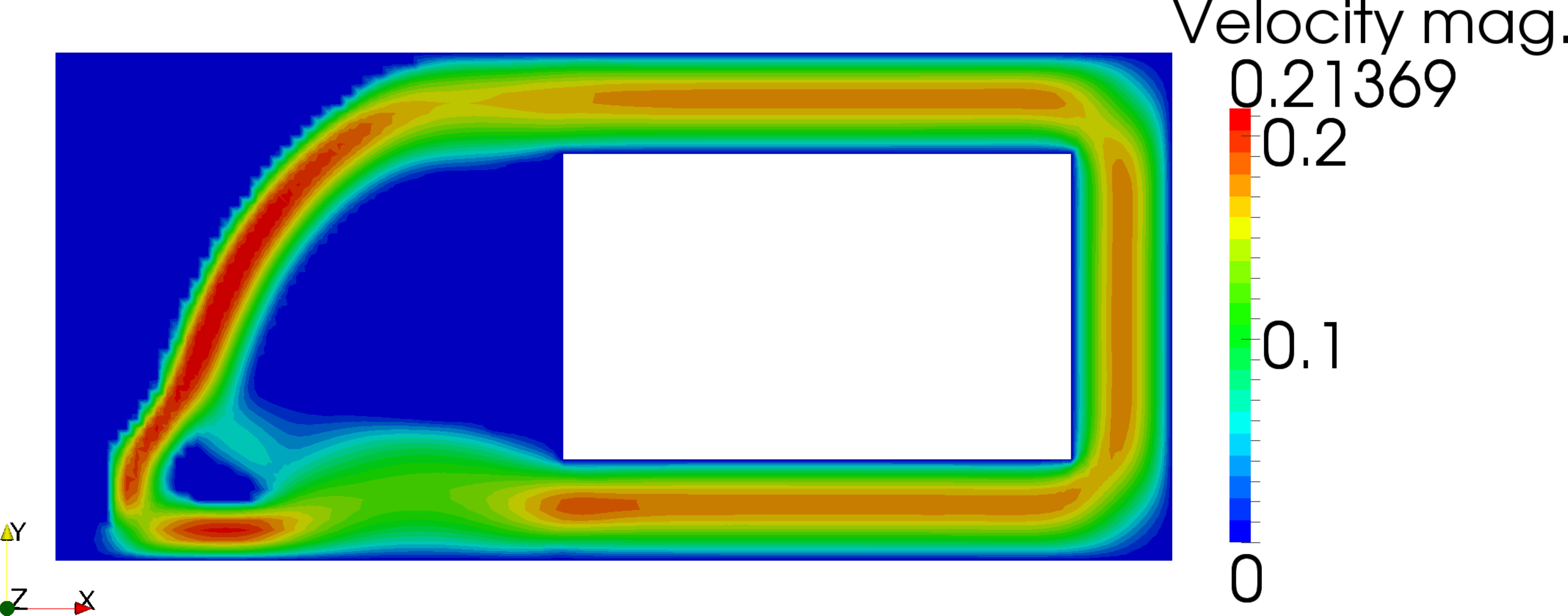}
\label{fig:cpunatconv2_contin0-c}}
\caption{Optimised design along with streamlines and corresponding temperature and velocity fields for clockwise massflow for combination 3 of the natural convection micropump. \\
Objective function: $f_{m\! f} = -2.350\cdot10^{-5}$ - Design iterations: 300} \label{fig:cpunatconv2_contin0}
\end{figure}
Figure \ref{fig:cpunatconv2_contin0} shows the optimised design along with the temperature and velocity fields for the third combination listed in table \ref{tab:natconvpump_bcs}, where the problem domain has been tipped on its side. The gravitational direction remains in the negative $x_{2}$-directions, which means that the gravitational direction now becomes orthogonal to the longest dimension of the domain. A significantly different design is obtained than for combination 1, which has the same boundary conditions but a different orientation. Here it is again seen to be beneficial to include a small low conductivity gap in order to provide a long hot wall bordering the flow channel and thus increase the natural convection effect. However, for this combination of boundary conditions the relative densities are 1 in this region, due to the convexity factor being set to $q_{f}=0$. It is interesting to note that the maximum velocity is much smaller than for combination 1. This is likely due to the fact that the total length of vertical channel, where the upwards and downwards action of buoyancy and gravity is in full effect, is smaller.

The presented results are mainly intended as a proof-of-concept to show that topology optimisation is a viable approach to design micropumps based on natural convection effects. However, the results were not easy to obtain with the current methodology and these difficulties will be further discussed in section \ref{sec:discconc}.


\section{Discussion and conclusion} \label{sec:discconc}
\vspace{-2pt}

This study shows that topology optimisation is a viable approach for designing heat sink geometries cooled by natural convection and micropumps powered by natural convection. The examples highlight that natural convection can be exploited for both convective cooling effects as well as for generating fluid motion.

However, there are several difficulties when dealing with the natural convection problems treated in this paper and the main difficulties have been with solving the system of non-linear equations.
The underlying problems are highly non-linear and highly coupled, posing great difficulties for the Newton solver. The results presented in this paper were all obtained using a damped Newton solver using a constant damping factor throughout the non-linear iterations as well as the optimisation iterations. The reason for why this rather simple, but ineffective, scheme was chosen is that it provides a very robust solver. The lack of intelligently designed and/or heuristic update criteria for the damping factor means that the next solution update will never be rejected. This provides robustness in that the non-linear solver is less prone to get stuck at local stationary points in the solution space as it will simply move past them, whereas many line search methods may get stuck due to the requirement of a reduction in some convergence measure. However, using a constant damping factor less than unity does not take advantage of the full quadratic convergence of Newton's method when near the solution and yields an extremely ineffective solver, especially as the damping factor is chosen on the conservative side to ensure convergence throughout the optimisation procedure. This is not a significant problem when one only wishes to solve the state problem a single time, but in topology optimisation the state problem must be solved several hundred times in succession and an efficient non-linear solution method would therefore be desirable. Various methods that attempt to globalise Newton's method have been tested, but have not shown promise for the natural convection problems treated in this paper. Further investigation into the development of robust update criteria are left as future research.

Another problem, related to the non-linearity of the underlying equations, is encountered with the assumption of laminar steady-state flow. This is a reasonable assumption which simplifies the calculations significantly, but it is impossible to ensure that the problem will remain steady throughout the optimisation process. The problem is especially prevalent for highly convective cases, where it is very easy for the optimisation procedure to go through intermediate designs that trigger unsteady effects in the flow, leading to oscillatory behaviour or divergence of the Newton solver. This places a quite severe restriction on the application of the presented methodology, as is always the case when one makes assumptions, in that one cannot a priori guarantee that the assumptions are met. It is quite difficult to find values for the critical Grashof numbers in the literature, especially for complex geometries. However, as qualitative guidance one can make use of the limits predicted for e.g. cylinders \cite{Priye2012}, infinite plates \cite{Chandrasekhar1961} or shallow porous cavities \cite{Vasseur1989}. 

Applying the presented methodology to the design of natural convection micropumps has been successful, but certainly not without difficulties. When working with an objective functional based on the fluid velocity field, such as maximising the volumetric flow, a new problem arises that cannot be circumvented with certainty using penalisation of the impermeability and the thermal conductivity.
For natural convection around and inside a porous medium, the relationship between the effective impermeability and the magnitudes of the velocities inside of the porous material does not seem to be straight forward. For forced convective flow, a monotonous decrease is experienced in the velocities inside of the porous material, with respect to increasing the maximum impermeability. However, it has been observed that intermediate relative densities can be favoured by the optimisation when using the standard penalisation of the impermeability and the thermal conductivity for natural convection flow problems.
It appears that having an intermediate impermeability in large parts of the design domain produces an amplification of the natural convection effect, thereby increasing the velocity of the fluid flow. Parallels can be drawn to the studies carried out by Vasseur et al. \cite{Vasseur1989} and Lauriat et al. \cite{Lauriat1987} which indicate that the presence of a porous medium can yield higher flow velocities in cavities subjected to natural convection, than for cavities filled only with fluid. This indicates that introducing a fictitious porous medium to facilitate topology optimisation of natural convection fluid flow problems is highly non-trivial and needs to be investigated further, especially for the cases considered here where the differences in thermal conductivity of the fluid and solid/porous phases are taken into account. It is possible that these problems could be solved by making the Boussinesq forcing term design-dependent as is done with the body force driven flows considered by Deng et al. \cite{Deng2012a}. However, this is outside the scope of the current study and is left as a subject for future research.

Generally when applying the density-based topology optimisation method to multiphysics problems, where one needs to interpolate an increasing number of physical properties, the selection of penalisation/convexity parameters becomes increasingly difficult and non-trivial. For thermal compliance problems, like the heat sink problem presented in this paper, good results have been achieved by having a constant high convexity of the impermeability interpolation function while performing a continuation approach on the convexity parameter for the effective conductivity interpolation. A possibility for future research could be to consider physical homogenisation-based interpolation.

The presented methodology is actually already implemented for three-dimensional calculation and optimisation, but has been restricted to two-dimensions in this initial paper in order to investigate and tune the application of the density-based topology optimisation approach for small and plane natural convection problems. The extension of the methodology to three-dimensional problems is thus trivial, however, the huge increase in computational work for three-dimensional problems has been the major obstacle. A paper presenting applications to large scale three-dimensional problems is under preparation.

\vspace{-6pt}

\newpage
\ack The authors wish to thank Boyan S. Lazarov for stimulating discussions and help with regards to C++ and performing parallel calculations on the TopOpt-cluster.

\appendix
\section{Stabilisation parameters} \label{sec:stabpar}

In order to have an effective, and not overly diffusive, contribution from the stabilisation, the stabilisation parameters, $\tau_{*}$, need to be carefully determined in order to ensure that enough stabilisation is applied to ensure a non-oscillatory solution, but without adding too much and thereby sacrificing solution accuracy. Countless papers have been published on the subject of determining the best stabilisation parameter, see for example \cite{Tezduyar2003a} for an overview. 

The current implementation is based on the stabilisation parameters defined in \cite{Tezduyar2008} which are so-called UGN-based stabilisation parameters. The stabilisation parameters are defined as follows:
\begin{equation}
\tau_{PS} = \tau_{SU} = \left( \frac{1}{{\tau_{SUGN1}}^{r}} + \frac{1}{{\tau_{SUGN3}}^{r}} + {\alpha_{e}}^{r} \right)^{-1/r}
\end{equation}
which is an approximate min-function that switches between the two parameters:
\begin{align}
\tau_{SUGN1} &= \frac{h_{UGN}}{2 \lVert \mathbf{u}^{h} \rVert} \\
\tau_{SUGN3} &= \frac{{h_{RGN}}^2}{4 Pr}
\end{align}
and the inverse of the element Brinkman friction factor, $\alpha_{e}$\footnote{The stabilisation parameters are made dependent on the Brinkman coefficient of the elements, in order to mitigate problems observed at solid-fluid interfaces during the research work, where the pressure distribution showed oscillatory behaviour and exhibited large variations at the solid-fluid interfaces. Inspired by the papers by Masud \cite{Masud2007} and Braack et al. \cite{Braack2007}, including the element Brinkman coefficient makes sure that the stabilisation takes the reaction/porosity-dominance into account when the inverse permeability is large (solid and intermediate regions).}, based on a switching parameter, \textit{r}, which is set to 2. $\tau_{SUGN2}$ is the stabilisation factor corresponding to transient effects and is thus not included here due to the steady-state formulation.

The lengthscales used in the above parameters are defined as:
\begin{align}
h_{UGN} &= 2 \lVert \mathbf{u}^{h} \rVert \left( \sum_{a=1}^{n_{en}}\lvert \mathbf{u}^{h}\cdot \boldsymbol{\nabla} N_{ua} \rvert \right)^{-1} \\
h_{RGN} &= 2\left( \sum_{a=1}^{n_{en}}\lVert\mathbf{r}\cdot\boldsymbol{\nabla} N_{ua}\rVert  \right)^{-1}
\end{align}
where $n_{en}$ is the number of nodes per element, $N_{ua}$ is the velocity shape function for node $a$ and $\vecr{r}$ is a unit vector defined in the velocity-gradient direction:
\begin{equation}
\vecr{r} = \frac{\boldsymbol{\nabla}\lvert \vecr{u}^{h}\rvert}{ \lVert \boldsymbol{\nabla}\lvert\vecr{u}^{h}\rvert\rVert}
\end{equation}

The stabilisation parameters are taken to be constant within each element, so the expressions above are evaluated at the element centres. This is an approximation to the full stabilisation parameters based on element matrices and vectors which has been simplified based on the assumption of using a single integration point to evaluate the stabilisation parameters \cite{Tezduyar2003}. All norms in this section are defined to be the 2-norm.

Likewise, for the thermal SUPG stabilisation:
\begin{equation}
\tau_{SU_{T}} = \left( \frac{1}{{\tau_{SUGN1_{T}}}^{r}} + \frac{1}{{\tau_{SUGN3_{T}}}^{r}} \right)^{-1/r}
\end{equation}
where:
\begin{align}
\tau_{SUGN1_{T}} &= \tau_{SUGN1} \\
\tau_{SUGN3_{T}} &= \frac{{h_{RGN_{T}}}^2}{4 K_{e}}
\end{align}
The length-scale used above is defined as: 
\begin{equation}
h_{RGN_{T}} = 2 \left( \sum_{a=1}^{n_{en}}\lvert\vecr{r}_{T}\cdot\boldsymbol{\nabla} N_{ta}\lvert \right)^{-1}
\end{equation}
where $n_{en}$ is the number of nodes per element, $N_{ta}$ is the temperature shape function for node $a$ and $\vecr{r}_{T}$ is a unit vector defined in the temperature-gradient direction:
\begin{equation}
\vecr{r}_{T} = \frac{\boldsymbol{\nabla} {T}^{h}}{\lVert\boldsymbol{\nabla} {T}^{h}\lVert}
\end{equation}

\bibliography{bib}{}

\begin{thebibliography}{10}
\providecommand{\url}[1]{\texttt{#1}}
\providecommand{\urlprefix}{URL }
\expandafter\ifx\csname urlstyle\endcsname\relax
  \providecommand{\doi}[1]{doi:\discretionary{}{}{}#1}\else
  \providecommand{\doi}{doi:\discretionary{}{}{}\begingroup
  \urlstyle{rm}\Url}\fi

\bibitem{Bar-Cohen2003}
Bar-Cohen A, Watwe AA, Prasher RS. \emph{Heat Transfer Handbook}, chap.~13.
  John Wiley \& Sons, 2003. ISBN: 978-0-471-39015-2.

\bibitem{Krishnan2004}
Krishnan M, Agrawal N, Burns MA, Ugaz VM. Reactions and fluidics in
  miniaturized natural convection systems. \emph{Analytical Chemistry}  2004;
  \textbf{76}(21):6254--6265, \doi{10.1021/ac049323u}.

\bibitem{Morrison1992}
Morrison AT. Optimization of heat sink fin geometries for heat sinks in natural
  convection. \emph{InterSociety Conference on Thermal Phenomena in Electronic
  Systems, I-THERM III}, 1992, \doi{10.1109/ITHERM.1992.187753}.

\bibitem{Bahadur2005}
Bahadur R, Bar-Cohen A. Thermal design and optimization of natural convection
  polymer pin fin heat sinks. \emph{IEEE Transactions on Components and
  Packaging Technologies}  2005; \textbf{28}(2):238--246,
  \doi{10.1109/TCAPT.2005.848498}.

\bibitem{Muddu2011}
Muddu R, Hassan YA, Ugaz VM. Chaotically accelerated polymerase chain reaction
  by microscale {R}ayleigh-{B}\'enard convection. \emph{Angewandte Chemie
  International Edition}  2011; \textbf{50}(13):3048--3052,
  \doi{10.1002/anie.201004217}.

\bibitem{Bendsoee1988}
{Bends\o e} MP, Kikuchi N. Generating optimal topologies in structural design
  using a homogenization method. \emph{Computer Methods in Applied Mechanics
  and Engineering}  1988; \textbf{71}(2):197--224,
  \doi{10.1016/0045-7825(88)90086-2}.

\bibitem{Bendsoee1989}
{Bends\o e} MP. Optimal shape design as a material distribution problem.
  \emph{Structural optimization}  1989; \textbf{1}(4):193--202,
  \doi{10.1007/BF01650949}.

\bibitem{Zhou1991}
Zhou M, Rozvany GIN. The {COC} algorithm, part {II}: Topological, geometrical
  and generalized shape optimization. \emph{Computer Methods in Applied
  Mechanics and Engineering}  1991; \textbf{89}(1-3):309--336,
  \doi{10.1016/0045-7825(91)90046-9}.

\bibitem{Rozvany1992}
Rozvany GIN, Zhou M, Birker T. Generalized shape optimization without
  homogenization. \emph{Structural Optimization}  1992;
  \textbf{4}(3-4):250--252, \doi{10.1007/BF01742754}.

\bibitem{Duhring2008}
Duhring MB, Jensen JS, Sigmund O. Acoustic design by topology optimization.
  \emph{Journal of Sound and Vibration}  2008; \textbf{317}(3-5):557--575,
  \doi{10.1016/j.jsv.2008.03.042}.

\bibitem{Jensen2011}
Jensen JS, Sigmund O. Topology optimization for nano-photonics. \emph{Laser and
  Photonics Reviews}  2011; \textbf{5}(2):308--321,
  \doi{10.1002/lpor.201000014}.

\bibitem{Borrvall2003}
Borrvall T, Petersson J. Topology optimization of fluids in {S}tokes flow.
  \emph{International Journal for Numerical Methods in Fluids}  2003;
  \textbf{41}(1):77--107, \doi{10.1002/fld.426}.

\bibitem{Bendsoee2003}
{Bends\o e} MP, Sigmund O. \emph{Topology Optimization: Theory, Methods and
  Applications}. Springer, 2003. ISBN: 3-540-42992-1.

\bibitem{Gersborg-Hansen2006}
Gersborg-Hansen A, e MPB, Sigmund O. Topology optimization of heat conduction
  problems using the finite volume method. \emph{Structural Multidisciplinary
  Optimization}  2006; \textbf{31}(4):251--259,
  \doi{10.1007/s00158-005-0584-3}.

\bibitem{Sigmund2001a}
Sigmund O. Design of multiphysics actuators using topology optimization - part
  {I}: One-material structures. \emph{Computer Methods in Applied Mechanics and
  Engineering}  2001; \textbf{190}(49-50):6577--6604,
  \doi{10.1016/S0045-7825(01)00251-1}.

\bibitem{Yin2002}
Yin L, Ananthasuresh G. A novel topology design scheme for the multi-physics
  problems of electro-thermally actuated compliant micromechanisms.
  \emph{Sensors and Actuators}  2002; \textbf{97-98}:599--609,
  \doi{10.1016/S0924-4247(01)00853-6}.

\bibitem{Yoon2005}
Yoon GH, Kim YY. The element connectivity parameterization formulation for the
  topology design optimization of multiphysics systems. \emph{International
  Journal for Numerical Methods in Engineering}  2005;
  \textbf{64}(12):1649--1677, \doi{10.1002/nme.1422}.

\bibitem{Bruns2007}
Bruns T. Topology optimization of convection-dominated, steady-state heat
  transfer problems. \emph{International Journal of Heat and Mass Transfer}
  2007; \textbf{50}(15-16):2859--2873,
  \doi{10.1016/j.ijheatmasstransfer.2007.01.039}.

\bibitem{Iga2009}
Iga A, Nishiwaki S, Izui K, Yoshimura M. Topology optimization for thermal
  conductors considering design-dependent effects, including heat conduction
  and convection. \emph{International Journal of Heat and Mass Transfer}  2009;
  \textbf{52}(11-12):2721--2732,
  \doi{10.1016/j.ijheatmasstransfer.2008.12.013}.

\bibitem{Ahn2010a}
Ahn SH, Cho S. Level set-based topological shape optimization of heat
  conduction problems considering design-dependent convection boundary.
  \emph{Numerical Heat Transfer, Part B: Fundamentals}  2010;
  \textbf{58:5}(5):304--322, \doi{10.1080/10407790.2010.522869}.

\bibitem{Gersborg-Hansen2005}
Gersborg-Hansen A, Sigmund O, Haber R. Topology optimization of channel flow
  problems. \emph{Structural Multidisciplinary Optimization}  2005;
  \textbf{30}(3):181--192, \doi{10.1007/s00158-004-0508-7}.

\bibitem{Olesen2006}
Olesen LH, Okkels F, Bruus H. A high-level programming-language implementation
  of topology optimization applied to steady-state {N}avier-{S}tokes flow.
  \emph{International Journal for Numerical Methods in Engineering}  2006;
  \textbf{65}(7):957--1001, \doi{10.1002/nme.1468}.

\bibitem{Andreasen2009}
Andreasen CS, Gersborg AR, Sigmund O. Topology optimization of microfluidic
  mixers. \emph{International Journal for Numerical Methods in Fluids}  2009;
  \textbf{61}(5):498--513, \doi{10.1002/fld.1964}.

\bibitem{Okkels2007}
Okkels F, Bruus H. Scaling behavior of optimally structured catalytic
  microfluidic reactors. \emph{Phys. Rev. E}  2007; \textbf{75}(1):016\,301,
  \doi{10.1103/PhysRevE.75.016301}.

\bibitem{Deng2011}
Deng Y, Liu Z, Zhang P, Liu Y, Wu Y. Topology optimization of unsteady
  incompressible {N}avier-{S}tokes flows. \emph{Journal of Computational
  Physics}  2011; \textbf{230}(17):6688--6708, \doi{10.1016/j.jcp.2011.05.004}.

\bibitem{Kreissl2011}
Kreissl S, Pingen G, Maute K. Topology optimization for unsteady flow.
  \emph{International Journal for Numerical Methods in Engineering}  2011;
  \textbf{87}(13):1229--1253, \doi{10.1002/nme.3151}.

\bibitem{Yoon2010a}
Yoon GH. Topology optimization for stationary fluid-structure interaction
  problems using a new monolithic formulation. \emph{International Journal for
  Numerical Methods in Engineering}  2010; \textbf{82}(5):591--616,
  \doi{10.1002/nme.2777}.

\bibitem{Deng2012a}
Deng Y, Liu Z, Wu Y. Topology optimization of steady and unsteady
  incompressible {N}avier-{S}tokes flows driven by body forces.
  \emph{Structural Multidisciplinary Optimization}  2013;
  \textbf{47}(4):555--570, \doi{10.1007/s00158-012-0847-8}.

\bibitem{Guest2006}
Guest JK, Prevost JH. Topology optimization of creeping fluid flows using a
  {D}arcy-{S}tokes finite element. \emph{International Journal for Numerical
  Methods in Engineering}  2006; \textbf{66}(3):461--484,
  \doi{10.1002/nme.1560}.

\bibitem{Zhou2008}
Zhou S, Li Q. A variational level set method for the topology optimization of
  steady-state {N}avier-{S}tokes flow. \emph{Journal of Computational Physics}
  2008; \textbf{227}(24):10\,178--10\,195, \doi{10.1016/j.jcp.2008.08.022}.

\bibitem{Challis2009}
Challis VJ, Guest JK. Level set topology optimization of fluids in {S}tokes
  flow. \emph{International Journal for Numerical Methods in Engineering}
  2009; \textbf{79}(10):1284--1308, \doi{10.1002/nme.2616}.

\bibitem{Kreissl2012}
Kreissl S, Maute K. Levelset based fluid topology optimization using the
  extended finite element method. \emph{Structural Multidisciplinary
  Optimization}  2012; \textbf{46}(3):311--326,
  \doi{10.1007/s00158-012-0782-8}.

\bibitem{Kontoleontos2012}
Kontoleontos EA, Papoutsis-Kiachagias EM, Zymaris AS, Papadimitriou DI,
  Giannakoglou KC. Adjoint-based constrained topology optimization for viscous
  flows, including heat transfer. \emph{Engineering Optimization}  2013;
  \textbf{45}(8):941--961, \doi{10.1080/0305215X.2012.717074}.

\bibitem{Matsumori2013}
Matsumori T, Kondoh T, Kawamoto A, Nomura T. Topology optimization for
  fluid-thermal interaction problems under constant input power.
  \emph{Structural Multidisciplinary Optimization}  2013;
  \textbf{47}(4):571--581, \doi{10.1007/s00158-013-0887-8}.

\bibitem{Yoon2010}
Yoon GH. Topological design of heat dissipating structure with forced
  convective heat transfer. \emph{Journal of Mechanical Science and Technology}
   2010; \textbf{24}(6):1225--1233, \doi{10.1007/s12206-010-0328-1}.

\bibitem{Dede2010}
Dede EM. Multiphysics optimization, synthesis, and application of jet
  impingement target surfaces. \emph{The 12th IEEE Intersociety Conference on
  Thermal and Thermomechanical Phenomena in Electronic Systems (ITherm)}, 2010,
  \doi{10.1109/ITHERM.2010.5501408}.

\bibitem{Lee2012}
Lee K. Topology optimization of convective cooling system designs. Ph{D}
  {T}hesis, University of Michigan 2012.
  \urlprefix\url{http://deepblue.lib.umich.edu}.

\bibitem{McConnell2012}
McConnell C, Pingen G. Multi-layer, pseudo 3{D} thermal topology optimization
  of heat sinks. \emph{Proceedings of the ASME 2012 International Mechanical
  Engineering Congress \& Exposition}, IMECE: Houston, Texas, USA, 2012.

\bibitem{Marck2013}
Marck G, Nemer M, Harion JL. Topology optimization of heat and mass transfer
  problems: Laminar flow. \emph{Numerical Heat Transfer, Part B: Fundamentals}
  2013; \textbf{63}(6):508--539, \doi{10.1080/10407790.2013.772001}.

\bibitem{Koga2013}
Koga AA, Lopes ECC, Nova HFV, de~Lima CR, Silva ECN. Development of heat sink
  device by using topology optimization. \emph{International Journal of Heat
  and Mass Transfer}  2013; \textbf{64}:759--772,
  \doi{10.1016/j.ijheatmasstransfer.2013.05.007}.

\bibitem{Priye2012}
Priye A, Hassan YA, Ugaz VM. Education: {DNA} replication using microscale
  natural convection. \emph{Lab on a Chip}  2012; \textbf{12}:4946--4954,
  \doi{10.1039/c2lc40760d}.

\bibitem{Hughes1986a}
Hughes TJ, France LP, Balestra M. A new finite element formulation for
  computational fluid dynamics {V} - circumventing the {B}abuska-{Brezzi}
  condition: a stable {P}etrov-{G}alerkin formulation of the {S}tokes problem
  accomodating equal-order interpolations. \emph{Computer Methods in Applied
  Mechanics and Engineering}  1986; \textbf{59}(1):85--99,
  \doi{10.1016/0045-7825(86)90025-3}.

\bibitem{Tezduyar1992a}
Tezduyar TE, Mittal S, Ray S, Shih R. Incompressible flow computations with
  stabilized bilinear and linear equal-order-interpolation velocity-pressure
  elements. \emph{Computer Methods in Applied Mechanics and Engineering}  1992;
  \textbf{95}(2):221--242, \doi{10.1016/0045-7825(92)90141-6}.

\bibitem{Brooks1982}
Brooks AN, Hughes TJ. Streamline {U}pwind/{P}etrov-{G}alerkin formulations for
  convection dominated flows with particular emphasis on the incompressible
  {N}avier-{S}tokes equations. \emph{Computer Methods in Applied Mechanics and
  Engineering}  1982; \textbf{32}(1-3):199--259,
  \doi{10.1016/0045-7825(82)90071-8}.

\bibitem{Stolpe2001a}
Stolpe M, Svanberg K. An alternative interpolation scheme for minimum
  compliance topology optimization. \emph{Structural Multidisciplinary
  Optimization}  2001; \textbf{22}(2):116--124, \doi{10.1007/s001580100129}.

\bibitem{Michaleris1994}
Michaleris P, Tortorelli DA, Vidal CA. Tangent operators and design sensitivity
  formulations for transient non-linear coupled problems with applications to
  elastoplasticity. \emph{International Journal for Numerical Methods in
  Engineering}  1994; \textbf{37}(14):2471--2499, \doi{10.1002/nme.1620371408}.

\bibitem{Sigmund1998}
Sigmund O, Petersson J. Numerical instabilities in topology optimization: A
  survey on procedures dealing with checkerboards, mesh-dependencies and local
  minima. \emph{Structural Optimization}  1998; \textbf{16}(1):68--75,
  \doi{10.1007/BF01214002}.

\bibitem{Alexandersen2013}
Alexandersen J. Topology optimisation for coupled convection problems. Master's
  {T}hesis, Technical University of Denmark 2013.
  \urlprefix\url{http://orbit.dtu.dk}.

\bibitem{Bruns2001}
Bruns TE, Tortorelli DA. Topology optimization of non-linear elastic structures
  and compliant mechanisms. \emph{Computer Methods in Applied Mechanics and
  Engineering}  2001; \textbf{190}(26-27):3443--3459,
  \doi{10.1016/S0045-7825(00)00278-4}.

\bibitem{Bourdin2001}
Bourdin B. Filters in topology optimization. \emph{International Journal for
  Numerical Methods in Engineering}  2001; \textbf{50}(9):2143--2158,
  \doi{10.1002/nme.116}.

\bibitem{Sigmund2007}
Sigmund O. Morphology-based black and white filters for topology optimization.
  \emph{Structural Multidisciplinary Optimization}  2007;
  \textbf{33}(4-5):401--424, \doi{10.1007/s00158-006-0087-x}.

\bibitem{Wang2010}
Wang F, Lazarov BS, Sigmund O. On projection methods, convergence and robust
  formulations in topology optimization. \emph{Structural Multidisciplinary
  Optimization}  2011; \textbf{43}(6):767--784,
  \doi{10.1007/s00158-010-0602-y}.

\bibitem{Aage2013}
Aage N, Lazarov BS. Parallel framework for topology optimization using the
  method of moving asymptotes. \emph{Structural and Multidisciplinary
  Optimization}  2013; \textbf{47}(4):493--505,
  \doi{10.1007/s00158-012-0869-2}.

\bibitem{Stroustrup}
Stroustrup B. \emph{The C++ Programming Language}. Addison-Wesley. ISBN:
  0-201-88954-4.

\bibitem{Amestoy2000}
Amestoy PR, Duff IS, L'Excellent JY. Multifrontal parallel distributed
  symmetric and unsymmetric solvers. \emph{Computer Methods in Applied
  Mechanics and Engineering}  2000; \textbf{184}(2-4):501--520,
  \doi{10.1016/S0045-7825(99)00242-X}.

\bibitem{Svanberg1987}
Svanberg K. The method of moving asymptotes - a new method for structural
  optimization. \emph{International Journal for Numerical Methods in
  Engineering}  1987; \textbf{24}(2):359--373, \doi{10.1002/nme.1620240207}.

\bibitem{Chandrasekhar1961}
Chandrasekhar S. \emph{Hydrodynamic and Hydromagnetic Stability}. Clarendon
  Press, 1961.

\bibitem{Vasseur1989}
Vasseur P, Wang CH, Sen M. The {B}rinkman model for natural convection in a
  shallow porous cavity with uniform heat flux. \emph{Numerical Heat Transfer,
  Part A: Applications}  1989; \textbf{15}(2):221--242,
  \doi{10.1080/10407788908944686}.

\bibitem{Lauriat1987}
Lauriat G, Prased V. Natural convection in a vertical porous cavity: a
  numerical study for {B}rinkman-extended {D}arcy formulation. \emph{Journal of
  Heat Transfer - Transactions of the ASME}  1987; \textbf{109}:688--696,
  \doi{10.1115/1.3248143}.

\bibitem{Tezduyar2003a}
Tezduyar TE, Sathe S. Stabilisation parameters in {SUPG} and {PSPG}
  formulations. \emph{Journal of Computational and Applied Mechanics}  2003;
  \textbf{4}(1):71--88.

\bibitem{Tezduyar2008}
Tezduyar TE, Ramakrishnan S, Sathe S. Stabilized formulations for
  incompressible flows with thermal coupling. \emph{International Journal for
  Numerical Methods in Fluids}  2008; \textbf{57}(9):1189--1209,
  \doi{10.1002/fld.1743}.

\bibitem{Masud2007}
Masud A. A stabilized mixed finite element method for {D}arcy-{S}tokes flow.
  \emph{International Journal for Numerical Methods in Fluids}  2007;
  \textbf{54}(6-8):665--681, \doi{10.1002/fld.1508}.

\bibitem{Braack2007}
Braack M, Burman E, John V, Lube G. Stabilized finite element methods for the
  generalized {O}seen problem. \emph{Computer Methods in Applied Mechanics and
  Engineering}  2007; \textbf{196}(4-6):853--866,
  \doi{10.1016/j.cma.2006.07.011}.

\bibitem{Tezduyar2003}
Tezduyar TE. Computation of moving boundaries and interfaces and stabilization
  parameters. \emph{International Journal for Numerical Methods in Fluids}
  2003; \textbf{43}(5):555--575, \doi{10.1002/fld.505}.

\end{thebibliography}
\bibliographystyle{wileyj}
\end{document}